\documentclass{aa}  

\usepackage{graphicx}
\usepackage{txfonts}
\usepackage{subcaption}         
\usepackage{lscape}             
\usepackage{placeins}           

\usepackage{silence}
\usepackage{physics}

\usepackage{dcolumn}
    \newcolumntype{d}[1]{D{.}{\cdot}{#1}}
    \newcolumntype{.}{D{.}{.}{-2}}
    \newcolumntype{,}{D{,}{,}{-3}}

\newbox\tablebox    \newdimen\tablewidth \def\leaderfil{\leaders\hbox to 5pt{\hss.\hss}\hfil} \def\endPlancktable{\tablewidth=\columnwidth $$\hss\copy\tablebox\hss$$
    \vskip-\lastskip\vskip -2pt}
\def\endPlancktablewide{\tablewidth=\textwidth $$\hss\copy\tablebox\hss$$
    \vskip-\lastskip\vskip -2pt}
\def\doubleline{\vskip 3pt\hrule \vskip 1.5pt \hrule \vskip 5pt} \def\tablenote#1 #2\par{\begingroup \parindent=0.8em \abovedisplayshortskip=0pt\belowdisplayshortskip=0pt
    \noindent $$\hss\vbox{\hsize\tablewidth \hangindent=\parindent \hangafter=1 \noindent \hbox to \parindent{$^#1$\hss}\strut#2\strut\par}\hss$$ \endgroup}

\usepackage{xspace}  

\usepackage{tikz}
\usetikzlibrary{positioning, fit, calc, arrows.meta, backgrounds}

\definecolor{CFblue}{RGB}{60, 100, 165}
\definecolor{CForange}{RGB}{210, 135, 35}
\definecolor{CFgreen}{RGB}{75, 140, 80}

\usepackage[breaklinks,bookmarks=false,pdfpagemode=UseNone]{hyperref}

\usepackage{fontawesome5}
\providecommand{\orcid}[1]{\,\href{https://orcid.org/#1}{\textcolor[HTML]{A6CE39}{\faOrcid}}}

\newcommand{\cf}{\texttt{CosmoForge}\xspace}
\newcommand{\cc}{\texttt{CosmoCore}\xspace}
\newcommand{\qube}{\texttt{QUBE}\xspace}
\newcommand{\pics}{\texttt{PICSLike}\xspace}

\newcommand{\dvec}{\mathbf{d}}           
\newcommand{\svec}{\mathbf{s}}           
\newcommand{\nvec}{\mathbf{n}}           
\newcommand{\wvec}{\mathbf{w}}           
\newcommand{\Cmat}{\mathbf{C}}           
\newcommand{\Smat}{\mathbf{S}}           
\newcommand{\Nmat}{\mathbf{N}}           
\newcommand{\Neff}{\mathbf{N}_\mathrm{eff}}  
\newcommand{\Fmat}{\mathbf{F}}           
\newcommand{\Fbm}{\mathbf{F}^\mathrm{b}} 
\newcommand{\Amat}{\mathbf{A}}           
\newcommand{\Tmat}{\mathbf{T}}           
\newcommand{\Imat}{\mathbf{I}}           
\newcommand{\Vmat}{\mathbf{V}}           
\newcommand{\Lmat}{\boldsymbol{\Lambda}} 
\newcommand{\Kmat}{\mathbf{K}}           
\newcommand{\Mmat}{\mathbf{M}}           
\newcommand{\Pmat}{\mathbf{P}}           
\newcommand{\Emat}{\mathbf{E}}           
\newcommand{\Ctinv}{\widetilde{\Cmat}^{-1}}  
\newcommand{\nhat}{\hat{n}}              
\newcommand{\Wl}{W_\ell}                 
\newcommand{\Bl}{B_\ell}                 
\newcommand{\pl}{p_\ell}                 
\newcommand{\Cl}{C_\ell}                 
\newcommand{\Chat}{\hat{C}_\ell}         
\newcommand{\nmodes}{n_\mathrm{modes}}   
\newcommand{\npix}{N_\mathrm{pix}}       
\newcommand{\nside}{N_\mathrm{side}}     
\newcommand{\Nsim}{N_\mathrm{sims}}      
\newcommand{\nell}{n_\ell}               
\newcommand{\nspec}{n_s}                 
\newcommand{\lmin}{\ell_\mathrm{min}}                  
\newcommand{\lmax}{\ell_\mathrm{max}}                  
\newcommand{\lminsig}{\ell^\mathrm{Sig.}_\mathrm{min}} 
\newcommand{\lmaxsig}{\ell^\mathrm{Sig.}_\mathrm{max}} 
\newcommand{\fsky}{f_\mathrm{sky}}       
\newcommand{\nbins}{N_\mathrm{bins}}     
\newcommand{\dbell}{\Delta\ell}          
\newcommand{\Wmat}{\mathbf{W}}           
\newcommand{\params}{\boldsymbol{\theta}}
\newcommand{\Umat}{\mathbf{U}}           
\newcommand{\Ximat}{\boldsymbol{\Xi}}    

\begin{document}

\title{CosmoForge I}

\subtitle{A unified framework for QML power spectrum estimation and pixel-based likelihood analysis}

\titlerunning{CosmoForge I}
\authorrunning{Galloni \& Pagano}


\author{G. Galloni\orcid{0000-0002-2412-8311}\inst{1,2}\corrauth{giacomo.galloni@unife.it}
    \and L. Pagano\orcid{0000-0003-1820-5998}\inst{1,2,3}\email{luca.pagano@unife.it}
    }

\institute{Dipartimento di Fisica e Scienze della Terra, Universit\`a degli Studi di Ferrara, via Saragat 1, I-44122 Ferrara, Italy
\and Istituto Nazionale di Fisica Nucleare, Sezione di Ferrara, via Saragat 1, I-44122 Ferrara, Italy
\and Institut d'Astrophysique Spatiale, CNRS, Univ.\ Paris-Sud, Universit\'e Paris-Saclay, B\^at.\ 121, 91405 Orsay cedex, France}

\date{Draft \today}


\abstract
{Optimal power spectrum estimation on the largest angular scales of the cosmic microwave background relies on the Quadratic Maximum Likelihood (QML) estimator. Existing public implementations, however, each address only a subset of the problem and none combines power spectrum estimation with a self-consistent pixel-space likelihood within a single framework.}
{We present \cf, a public Python framework that unifies QML power spectrum estimation and pixel-based Gaussian likelihood evaluation for spin-0 and spin-2 fields on the sphere, with general (non-diagonal) noise covariances.}
{The framework is split into three installable packages: \cc (infrastructure), \qube (Fisher and QML estimation), and \pics (pixel-space likelihood). A common interface exposes two interchangeable computation bases --- a harmonic basis built on the Sherman--Morrison--Woodbury identity and a direct pixel-space basis --- selecting whichever is cheaper for the configuration at hand. Exact algorithmic optimisations reduce the Fisher cost to $\mathcal{O}(\lmax^4)$ for arbitrary noise covariances, with {\tt Numba} JIT compilation of the hot kernels and MPI parallelisation of the likelihood scan.}
{\cf reproduces the Planck low-$\ell$ Fortran reference implementation across both the QML and pixel-space likelihood pipelines, consistently with double-precision arithmetic. Native multipole binning and three output normalisations (deconvolved, decorrelated, window-convolved) are exposed through a single code path, and the same covariance infrastructure powers both QML estimation and likelihood evaluation.}
{\cf offers a general-purpose, modular, and validated tool for the optimal analysis of large-scale data on the sphere. It is publicly available, pip-installable, and extensible to non-CMB observables.}


\maketitle
\nolinenumbers


\section{Introduction}
The angular power spectrum of the Cosmic Microwave Background (CMB) is a cornerstone observable of modern Cosmology. Its measurement has driven precision constraints on the cosmological model, from the first detections of acoustic oscillations to the tight bounds on the tensor-to-scalar ratio $r$ from current experiments \citep{PlanckV2020, BICEPKeck2021, Tristram2022}. Next-generation experiments such as LiteBIRD \citep{LiteBIRD2023} and the Simons Observatory \citep{SimonsObservatory2019} will target large angular scales with unprecedented sensitivity, where the primordial gravitational wave signal in B-mode polarization is expected to peak and where the optical depth to reionization $\tau$ must be determined.

At these large angular scales, the limited number of available modes and the effects of partial sky coverage make optimal statistical methods essential. The Quadratic Maximum Likelihood (QML) estimator, introduced by \citet{Tegmark1997} and extended to polarization by \citet{TegmarkOliveiraCosta2001}, provides minimum-variance, unbiased power spectrum estimates by construction. Unlike pseudo-$C_\ell$ methods \citep{Hivon2002, Alonso2019}, which are computationally cheap but suboptimal on large scales, the QML approach accounts exactly for the mask-induced mode coupling and noise anisotropy through the full pixel-space covariance matrix. This optimality comes at a computational cost that scales as $\mathcal{O}(N_\mathrm{pix}^3)$ in the naive implementation, set by the inversion of the pixel-space covariance and by the dense matrix products in the Fisher-matrix trace, limiting its applicability to low-resolution maps ($N_\mathrm{side} \lesssim 64$). However, the large-scale science case --- primordial B-modes, reionization optical depth, parity violation, isotropy anomalies --- precisely requires this regime, making QML the method of choice for the most sensitive measurements planned in the coming decade. The QML estimator's optimality is not, however, restricted to the large-scale regime: at intermediate and small scales, where pseudo-$C_\ell$ methods are usually adopted for their cheap cost, QML can still deliver tighter error bars when the noise covariance is non-trivial or the mask geometry induces significant mode coupling that pseudo-$C_\ell$ approximations handle only asymptotically, opening room for deeper scientific targets where computational resources allow.

The pixel-space likelihood offers a complementary approach: rather than estimating the power spectrum as an intermediate step, it evaluates the full Gaussian log-likelihood $\ln \mathcal{L} = -\frac{1}{2}[\mathbf{d}^\top \mathbf{C}^{-1} \mathbf{d} + \ln|\mathbf{C}|]$ directly for each point in a cosmological parameter grid. This approach has been used extensively in low-$\ell$ CMB analyses, from the WMAP polarization likelihood \citep{Page2007} to the Planck low-$\ell$ pipelines \citep{PlanckXI2016}, and provides the ground truth against which approximate likelihood methods are validated \citep{Galloni2025}. Its combination with QML estimation in a single framework enables a self-consistent analysis pipeline from power spectrum recovery to parameter inference.

Over the past two decades, several QML implementations have been developed, each tailored to specific analysis needs: pixel-space polarization estimation \citep{Gruppuso2009}, the unified low-$\ell$ likelihood and QML formalism of \citet{Gjerlow2015}, which compared pixel, spherical-harmonic, and signal-to-noise eigenbasis representations on the same footing, cross-spectrum estimation for noise bias removal \citep{Vanneste2018}, optimised harmonic-space formulations \citep{BilbaoAhedo2021}, and scaling to large numbers of correlated fields for galaxy surveys \citep{Kvasiuk2025}. These codes have enabled important scientific results, but they individually address different subsets of the problem. Some are restricted to specific computation bases (pixel or harmonic, but not both), others assume diagonal noise covariance --- excluding the correlated noise patterns common in real CMB experiments, both space- and ground-based, and ubiquitous when residual systematics or component-separation outputs are propagated to the analysis covariance --- and none integrates a pixel-space likelihood module within the same framework. The situation echoes that of the pseudo-$C_\ell$ landscape before the development of NaMaster \citep{Alonso2019}, where well-validated codes such as MASTER \citep{Hivon2002}, PolSpice \citep{Chon2004}, Xpol \citep{Tristram2005}, and Xpure \citep{Grain2009} each handled specific cases well, but the community lacked a single public framework consolidating those capabilities into a maintained, validated tool.

In this paper, we present \textbf{\cf}, a unified Python framework for QML power spectrum estimation and pixel-based likelihood analysis of fields on the sphere. While designed with the CMB in mind, the framework applies to any signal characterised by an angular power spectrum (e.g.\ galaxy surveys, 21\,cm intensity mapping, gravitational-wave backgrounds) and to any combination of them. \cf supports the joint analysis of scalar (spin-0, denoted $S$) and tensor (spin-2, decomposed into gradient $G$ and curl $C$ components) fields with all auto- and cross-spectra ($SS, GG, CC, SG, SC, GC$; the CMB analogues are $TT, EE, BB, TE, TB, EB$) in arbitrary multi-field configurations, with automatic detection of field groups that decouple in both signal and noise for fast block-diagonal covariance handling. The same algorithm is exposed through two interchangeable computation bases --- a harmonic basis built on the Sherman--Morrison--Woodbury (SMW) identity \citep{PlanckXI2016} and a direct pixel-space basis with Cholesky inversion --- behind a common abstract interface that auto-selects the cheaper path. Native multipole binning is provided through a single code path that recovers per-multipole estimation as its $\Delta\ell = 1$ special case, with three output normalisations available without recomputation: deconvolved $\hat{C}_\ell = F^{-1} q$, decorrelated $F^{-1/2} q$ with unit covariance, and window-convolved estimates for direct comparison with theory. The same covariance infrastructure powers a pixel-space Gaussian likelihood over cosmological parameter grids, with MPI parallelism on the independent stages (Fisher trace pairs, independent simulations, parameter-grid points) and Numba JIT compilation of the hot kernels (Legendre polynomials, Wigner $d$-matrices, signal-matrix assembly), bringing the harmonic-basis Fisher to an exact $\mathcal{O}(\lmax^4)$ scaling for arbitrary noise covariances.

\cf is organised as three interconnected, independently pip-installable packages: \textbf{\cc} (computational foundation, including the computation-basis abstraction with its harmonic and pixel implementations and the SMW machinery), \textbf{\qube} (Quadratic maximum likelihood UnBiased Estimator: Fisher matrix and QML estimation), and \textbf{\pics} (Pixel-based Inference with Correlated-Skies Likelihood: pixel-space Gaussian likelihood over parameter grids). The full pipeline has been validated against the Planck low-$\ell$ Fortran reference of \citet{Pagano2020} and the {\tt simall} likelihood of \citet{PlanckV2020} to machine-precision tolerances (Sect.~\ref{sec:fortran_validation}). Forthcoming work will present specific features and scientific case studies.

The paper is organised as follows. Section~\ref{sec:formalism} reviews the QML formalism, the harmonic-space reformulation based on the Sherman--Morrison--Woodbury identity, native multipole binning, the three available output normalisations, and the pixel-space likelihood. Section~\ref{sec:architecture} describes the package architecture and its principal classes. Sections~\ref{sec:validation}--\ref{sec:performance} present the validation against the Planck low-$\ell$ reference, internal consistency and Monte Carlo tests, an end-to-end pixel-based inference demonstration on the tensor-to-scalar ratio, and the empirical performance characterisation. Section~\ref{sec:conclusions} summarises the work and outlines forthcoming developments. Appendices~\ref{app:spin_harmonics}--\ref{app:yaml} collect the spin-2 extension and the spin-weighted harmonic conventions, the matrix-determinant-lemma derivation of the log-determinant, the algorithmic optimisations of the harmonic-space implementation, the comparison of \cf with existing public QML codes, and the configuration interface together with the YAML reference.
\section{Formalism}
\label{sec:formalism}

In this section, we present the mathematical framework underlying the QML estimator and its implementation in \cf. We begin with the signal model and covariance structure (Sect.~\ref{sec:signal_model}), then present the QML estimator in pixel space (Sect.~\ref{sec:qml_estimator}). Section~\ref{sec:harmonic} introduces the harmonic-space reformulation based on the Sherman--Morrison--Woodbury identity, which provides an exact and computationally efficient alternative to direct pixel-space evaluation. For clarity, we present the formalism for spin-0 (scalar) fields throughout Sects.~\ref{sec:signal_model}--\ref{sec:harmonic}; the generalisation to spin-2 (tensor) fields is given in Appendix~\ref{app:spin_harmonics}. Sections~\ref{sec:binning}--\ref{sec:pixel_likelihood} describe multipole binning, the three normalisation modes, and the pixel-space likelihood, respectively.

\subsection{Signal model and covariance}
\label{sec:signal_model}

We consider a set of observations of one or more fields on the sphere, pixelised using the HEALPix scheme \citep{Gorski2005} at resolution $\nside$. After masking unreliable or contaminated pixels, the data vector $\dvec$ consists of the $\npix$ values at the retained pixel locations. In the simplest case of a single spin-0 field (e.g., CMB temperature, galaxy overdensity, or 21\,cm intensity), the data model reads

\begin{equation}
    \dvec = \svec + \nvec,
    \label{eq:data_model}
\end{equation}

where $\svec$ is the signal and $\nvec$ is the instrumental noise, both assumed to be zero-mean Gaussian random fields. Their statistical properties are fully characterised by their covariance matrices: the signal covariance $\Smat = \langle \svec\svec^\top \rangle$ and the noise covariance $\Nmat = \langle \nvec\nvec^\top \rangle$. The total data covariance is

\begin{equation}
    \Cmat = \Smat + \Nmat.
    \label{eq:total_cov}
\end{equation}

We emphasise that $\Nmat$ is a general symmetric positive-definite matrix: \cf does not require the noise to be diagonal (uncorrelated between pixels) or isotropic (uniform across the sky). This allows the treatment of realistic noise patterns, including $1/f$ noise present in both space-based and ground-based experiments, correlated residuals from component separation, and noise correlations between frequency channels in cross-correlation analyses.

The signal covariance encodes the angular power spectrum $\Cl$ of the underlying field. For a statistically isotropic signal observed through a circularly symmetric beam with transfer function $\Bl$ and pixel window function $\pl$, the elements of $\Smat$ are given by

\begin{equation}
    S_{pp'} = \sum_{\ell=\lminsig}^{\lmaxsig} \frac{2\ell+1}{4\pi}\, \Cl\, \Wl\, P_\ell(\cos\theta_{pp'}),
    \label{eq:signal_cov}
\end{equation}

where $P_\ell$ is the Legendre polynomial of degree $\ell$, $\theta_{pp'}$ is the angular separation between pixels $p$ and $p'$, and $\Wl = \Bl^2 \pl^2$ is the combined beam-and-pixel window function. The sum is truncated on both ends: $\lminsig \geq |s|$ since spin-$s$ harmonics vanish below $\ell = |s|$, with $\lminsig = 2$ required for spin-2 fields and conventional for CMB temperature analyses (where it also excludes the monopole and dipole; see Sect.~\ref{sec:configuration} for the per-component generalisation), and $\lmaxsig$ is set by the support of $\Wl$. \cf\ defaults to $\lmaxsig = 4\,\nside$, the largest multipole at which the HEALPix pixel window $\pl^2$ is tabulated by {\tt healpy} \citep{Gorski2005, Zonca2019}; this may be lowered whenever $\Wl$ is already suppressed below analysis precision at a smaller $\ell$ --- in particular when the beam $\Bl^2$ alone has killed the signal --- for a corresponding reduction in basis dimension.

The power spectrum estimation problem consists of recovering the set $\{\Cl\}$ from the data $\dvec$, given knowledge of $\Nmat$, the pixel geometry, and the beam.

\subsection{The QML estimator}
\label{sec:qml_estimator}

The Quadratic Maximum Likelihood (QML) estimator \citep{Tegmark1997, BondJaffeKnox1998} provides an optimal solution to this problem. It is defined through a two-step procedure: (i) compute a set of quadratic quantities $q_\ell$ from the data that are related to the angular power spectrum; (ii) normalise them using the Fisher information matrix $\Fmat$ to obtain unbiased estimates $\Chat$.

The Fisher information matrix quantifies the sensitivity of the likelihood to changes in the power spectrum. Its elements are
\begin{equation}
    F_{\ell\ell'} = \frac{1}{2}\,\tr\!\left[\Cmat^{-1} \Pmat_\ell\, \Cmat^{-1} \Pmat_{\ell'}\right],
    \label{eq:fisher}
\end{equation}
where $\Pmat_\ell \equiv \partial\Cmat/\partial \Cl = \partial\Smat/\partial \Cl$ is the derivative of the signal covariance with respect to the power spectrum at multipole $\ell$. From Eq.~\eqref{eq:signal_cov}, this is
\begin{equation}
    (\Pmat_\ell)_{pp'} = \frac{2\ell+1}{4\pi}\, \Wl\, P_\ell(\cos\theta_{pp'}).
    \label{eq:P_ell}
\end{equation}
The quadratic estimates are
\begin{equation}
    q_\ell = \frac{1}{2}\,\dvec^\top \Cmat^{-1} \Pmat_\ell\, \Cmat^{-1}\, \dvec - b_\ell,
    \label{eq:q_ell}
\end{equation}
where $b_\ell$ is the noise bias,
\begin{equation}
    b_\ell = \frac{1}{2}\,\tr\!\left[\Cmat^{-1}\Pmat_\ell\,\Cmat^{-1}\Nmat\right].
    \label{eq:noise_bias}
\end{equation}
The expectation value of $q_\ell$ satisfies $\langle q_\ell \rangle = \sum_{\ell'} F_{\ell\ell'}\, C_{\ell'}$, so that the QML power spectrum estimate
\begin{equation}
    \Chat = \sum_{\ell'} (\Fmat^{-1})_{\ell\ell'}\, q_{\ell'}
    \label{eq:qml_estimate}
\end{equation}
is unbiased: $\langle\Chat\rangle = \Cl$. Its covariance is
\begin{equation}
    \mathrm{Cov}(\hat{C}_\ell, \hat{C}_{\ell'}) = (\Fmat^{-1})_{\ell\ell'},
    \label{eq:qml_cov}
\end{equation}
which saturates the Cram\'er--Rao bound, making the QML estimator minimum-variance among all unbiased estimators. The connection between this single-step estimator and the maximum of the full pixel-space likelihood is recovered in Sect.~\ref{sec:pixel_likelihood}; iterative-QML extensions building on the same architecture are deferred to forthcoming work (Sect.~\ref{sec:conclusions}).

\paragraph{Cross-correlation.} When two independent data sets $\dvec_1$ and $\dvec_2$ are available (e.g., maps from different detectors or frequency channels), two approaches to cross-spectrum estimation are possible.

The first \citep[following][]{TegmarkOliveiraCosta2001, Gjerlow2015} is to treat the two data sets as components of a joint data vector $\dvec = (\dvec_1, \dvec_2)$ with total covariance
\begin{equation}
    \Cmat_\mathrm{joint} = \begin{pmatrix} \Smat + \Nmat_1 & \Smat + \Nmat_{12} \\ \Smat + \Nmat_{12}^\top & \Smat + \Nmat_2 \end{pmatrix},
    \label{eq:C_joint}
\end{equation}
where $\Nmat_{12}$ encodes any noise component shared between $\dvec_1$ and $\dvec_2$ (zero in the simplest case of statistically independent splits), and apply the standard QML estimator (Eqs.~\ref{eq:fisher}--\ref{eq:qml_estimate}) to the full system. This yields the minimum-variance, unbiased estimator for all auto- and cross-spectra simultaneously, at the cost of doubling the data dimension. This formulation naturally accommodates residual noise correlations between the two data sets: any non-zero $\Nmat_{12}$ enters the off-diagonal blocks and is folded into both the estimator and its variance.

The second is a cross-correlation variant \citep{Vanneste2018, PlanckXLVI2016} in which the two data sets are filtered independently:
\begin{equation}
    q_\ell^\times = \frac{1}{2}\,\dvec_1^\top \Cmat_1^{-1} \Pmat_\ell\, \Cmat_2^{-1}\, \dvec_2,
    \label{eq:q_cross}
\end{equation}
with mode-mixing matrix $W_{\ell\ell'}^\times = \frac{1}{2}\,\tr\!\left[\Cmat_1^{-1}\Pmat_\ell\,\Cmat_2^{-1}\Pmat_{\ell'}\right]$. With independent noise between the two data sets the estimator is naturally noise-bias-free (any known $\Nmat_{12}$ is subtracted analogously\footnote{Analytical $\Nmat_{12}$ subtraction in Eq.~\eqref{eq:q_cross} is not yet exposed in the public API; the full multi-field path (Eq.~\ref{eq:C_joint}) supports it through the off-diagonal blocks.}). The independent filtering by $\Cmat_1^{-1}$ and $\Cmat_2^{-1}$ discards the inter-map signal correlation in the off-diagonal blocks of Eq.~\eqref{eq:C_joint}, so the estimator is unbiased but in general not minimum variance; in exchange, errors in $\Nmat_1$ or $\Nmat_2$ propagate into the variance but not the bias \citep{PlanckXLVI2016}. \cf supports both approaches: the full multi-field QML, and the cross-QML as a computationally lighter alternative when noise-bias freedom is the priority.

\paragraph{Computational cost.} In pixel space, the dominant costs are: (i) computing $\Cmat^{-1}$, which scales as $\mathcal{O}(\npix^3)$; and (ii) evaluating the Fisher matrix elements (Eq.~\ref{eq:fisher}), each of which requires the trace of a product of two $\npix \times \npix$ matrices. For $\nell$ multipoles and $\nspec$ spectra, the Fisher matrix has $(\nspec \cdot \nell)^2$ elements, and the total cost of the pixel-space implementation scales as $\mathcal{O}(\nspec^2\, \nell^2\, \npix^2 + \npix^3)$. This rapidly becomes prohibitive for high-resolution analyses, motivating the harmonic-space reformulation described in the next section.

\subsection{Harmonic-space formulation}
\label{sec:harmonic}

An algebraically equivalent but computationally more efficient formulation of the QML estimator can be obtained by factoring the signal covariance through the spherical harmonic basis. The harmonic factorisation $\Smat = \Vmat^\top\Lmat\Vmat$ that underpins this reformulation goes back to \citet{Tegmark1997}, where it was introduced as a computational device for the pixel-space Fisher matrix; \citet{Gjerlow2015} placed it within a unified low-$\ell$ likelihood framework that treats the spherical-harmonic, pixel, and signal-to-noise eigenbasis representations on a common footing. Its combination with the Sherman--Morrison--Woodbury (SMW) identity \citep{Woodbury1950, GolubVanLoan2013} for efficient covariance inversion was introduced in the context of the Planck low-$\ell$ likelihood \citep{PlanckXI2016}. \cf builds on this lineage, extending the SMW-accelerated harmonic basis from likelihood evaluation to QML power spectrum estimation and binding the two to a common pixel-space alternative through a single computation-basis interface (Sect.~\ref{sec:harmonic_basis_code}).

\subsubsection{The harmonic operator}
\label{sec:V_operator}

The derivative matrix $\Pmat_\ell$ (Eq.~\ref{eq:P_ell}) factors via the addition theorem for Legendre polynomials,
\begin{equation}
    P_\ell(\cos\theta_{pp'}) = \frac{4\pi}{2\ell+1} \sum_{m=-\ell}^{\ell} Y_{\ell m}(\nhat_p)\, Y_{\ell m}^*(\nhat_{p'}),
    \label{eq:addition_theorem}
\end{equation}
where $Y_{\ell m}$ are the spherical harmonics and $\nhat_p$ denotes the unit vector towards pixel $p$. Substituting into Eq.~\eqref{eq:signal_cov} and collecting terms, the signal covariance becomes
\begin{equation}
    \Smat = \Vmat^\top \Lmat\, \Vmat,
    \label{eq:S_factored}
\end{equation}
where $\Vmat$ is the $(\nmodes \times \npix)$ harmonic operator whose rows are the real spherical harmonics evaluated at the active pixel locations, and $\Lmat$ is the $(\nmodes \times \nmodes)$ diagonal matrix with entries $\Cl\, \Wl$, each repeated $(2\ell+1)$ times for the $m$-degeneracy. Here $\nmodes = \sum_{\ell=\lminsig}^{\lmaxsig} (2\ell+1)$ is the total number of harmonic modes.

In \cf, $\Vmat$ is constructed using real spherical harmonics with the normalisation convention
\begin{equation}
    Y_{\ell m}^\mathrm{real}(\theta, \phi) = \sqrt{\frac{2\ell+1}{4\pi}}\, N_{\ell m}\, P_\ell^{|m|}(\cos\theta) \times
    \begin{cases}
        \sqrt{2}\cos(m\phi) & m > 0 \\
        1 & m = 0 \\
        \sqrt{2}\sin(|m|\phi) & m < 0
    \end{cases}
    \label{eq:real_ylm}
\end{equation}
where $N_{\ell m} = \sqrt{(\ell-|m|)!/(\ell+|m|)!}$ and $P_\ell^m$ is the associated Legendre function. With this convention, the $(2\ell+1)/(4\pi)$ normalisation factor is absorbed into the basis functions, so that $\Cl$ inputs are physical power spectra (as output by Boltzmann codes such as {\tt CAMB} or {\tt CLASS}) without any pre-multiplication.

The rows of $\Vmat$ are ordered by azimuthal quantum number $|m|$ rather than by $\ell$: first all modes with $m=0$ (one per $\ell$), then pairs of $\cos/\sin$ modes for $|m|=1, 2, \ldots, \lmaxsig$.

\subsubsection{The Sherman--Morrison--Woodbury identity}
\label{sec:smw}

With the factorisation $\Smat = \Vmat^\top \Lmat\, \Vmat$, the total covariance takes the form of a low-rank update to the noise matrix:
\begin{equation}
    \Cmat = \Nmat + \Vmat^\top \Lmat\, \Vmat.
    \label{eq:C_factored}
\end{equation}
This structure allows the application of the Sherman--Morrison--Woodbury (SMW) identity, which expresses the inverse of such a matrix as
\begin{equation}
    \Cmat^{-1} = \Nmat^{-1} - \Nmat^{-1}\Vmat^\top \Kmat^{-1}\Vmat\Nmat^{-1},
    \label{eq:smw}
\end{equation}
where we have defined the SMW kernel
\begin{equation}
    \Kmat \equiv \Lmat^{-1} + \Vmat\Nmat^{-1}\Vmat^\top.
    \label{eq:K_kernel}
\end{equation}
The matrix $\Kmat$ has dimension $\nmodes \times \nmodes$, which is typically much smaller than $\npix \times \npix$: for an analysis at $\lmax = 32$, $\nmodes = 1\,085$ while $\npix$ may be several thousand depending on the sky coverage. The inversion of $\Kmat$ therefore costs $\mathcal{O}(\nmodes^3)$ instead of $\mathcal{O}(\npix^3)$. The projected combination $\Vmat\Cmat^{-1}\Vmat^{\top}$ that appears throughout the QML algebra reduces, via Eq.~\eqref{eq:smw}, to a difference of two large matrices that can suffer catastrophic cancellation in the cosmic-variance-limited regime; \cf evaluates it through an algebraically equivalent but numerically stable rewrite, described in Sect.~\ref{app:smw_stable}.

The log-determinant of $\Cmat$, needed for likelihood evaluation (Sect.~\ref{sec:pixel_likelihood}), can be obtained via the matrix determinant lemma:
\begin{equation}
    \ln|\Cmat| = \ln|\Nmat| + \ln|\Lmat| + \ln|\Kmat|.
    \label{eq:logdet}
\end{equation}
Each term on the right-hand side involves matrices of dimension at most $\nmodes$, and $\ln|\Nmat|$ needs to be computed only once (it is independent of $\Cl$).

\subsubsection{Fisher matrix and QML in harmonic space}
\label{sec:fisher_harmonic}

To express the Fisher matrix and QML estimates in harmonic space, it is convenient to define the projected inverse covariance
\begin{equation}
    \Ctinv \equiv \Vmat\,\Cmat^{-1}\,\Vmat^\top.
    \label{eq:projected_inv}
\end{equation}
Using Eq.~\eqref{eq:smw}, this can be written as
\begin{equation}
    \Ctinv = \Mmat - \Mmat\,\Kmat^{-1}\,\Mmat,
    \label{eq:projected_inv_smw}
\end{equation}
where $\Mmat \equiv \Vmat\,\Nmat^{-1}\,\Vmat^\top$ is the noise-weighted Gramian of the harmonic operator. Both $\Mmat$ and $\Kmat^{-1}$ are $(\nmodes \times \nmodes)$ matrices that are precomputed once and stored.

The derivative $\Pmat_\ell$ factors as $\Pmat_\ell = \Vmat^\top \Emat_\ell\, \Vmat$, where $\Emat_\ell$ is a diagonal matrix with entries equal to $\Wl$ at positions corresponding to modes with angular degree $\ell$, and zero elsewhere. Substituting into the Fisher matrix (Eq.~\ref{eq:fisher}) and using the cyclic property of the trace:
\begin{equation}
    F_{\ell\ell'} = \frac{1}{2}\,\tr\!\left[\Ctinv\, \Emat_\ell\; \Ctinv\, \Emat_{\ell'}\right].
    \label{eq:fisher_harmonic}
\end{equation}
Since $\Emat_\ell$ is diagonal, the products $\Ctinv\Emat_\ell$ amount to selecting and scaling columns of $\Ctinv$. Moreover, $\Emat_\ell$ has only $(2\ell+1)$ nonzero entries (the modes belonging to multipole $\ell$), so the trace can be reduced to a double sum over these mode indices, avoiding any dense matrix product. With this sparse-trace evaluation, computing the full Fisher matrix scales as $\mathcal{O}(\lmax^4)$ rather than the naive $\mathcal{O}(\nell^2\, \nmodes^2)$ that one would obtain from dense trace evaluation, while remaining mathematically exact for arbitrary noise covariance. Implementation details and benchmarks are deferred to Appendix~\ref{app:performance}.

Similarly, define the weighted harmonic data vector
\begin{equation}
    \wvec = \Vmat\,\Cmat^{-1}\,\dvec.
    \label{eq:weighted_data}
\end{equation}
Using the SMW expression for $\Cmat^{-1}$ (Eq.~\ref{eq:smw}), this can be computed efficiently as
\begin{equation}
    \wvec = \Vmat\Nmat^{-1}\dvec - \Mmat\,\Kmat^{-1}\,\Vmat\Nmat^{-1}\dvec,
    \label{eq:weighted_data_smw}
\end{equation}
requiring only matrix-vector products with precomputed quantities. The QML quadratic estimate then becomes
\begin{equation}
    q_\ell = \frac{1}{2}\,\wvec^\top\, \Emat_\ell\, \wvec - b_\ell,
    \label{eq:q_harmonic}
\end{equation}
which, since $\Emat_\ell$ is diagonal, reduces to a weighted sum over the squared harmonic coefficients at multipole $\ell$. This is the harmonic-space form actually evaluated by the {\tt Spectra} class (Sect.~\ref{sec:qube_code}) when the harmonic basis is selected.

\subsubsection{Multipole switching}
\label{sec:ell_switching}

In many analyses, the power spectrum is varied only over a limited range of multipoles $\ell \in [\lmin, \lmax]$ --- the \emph{inference window} --- while the signal at the multipoles bracketing it on either side is held fixed at a fiducial model: low-$\ell$ multipoles $\ell \in [\lminsig, \lmin)$ on the lower end, and high-$\ell$ multipoles $\ell \in (\lmax, \lmaxsig]$ on the upper end. This situation arises naturally when the science target is at large angular scales (e.g., primordial B-modes at $\ell \lesssim 30$) but the signal extends to smaller scales through the beam; symmetrically, it allows low-multipole content (e.g., the CMB temperature monopole and dipole) to be held at a fiducial value while still entering the model covariance, when the inference window starts at $\lmin = 2$.

Following the approach of \citet{PlanckXI2016}, the fixed-multipole contribution can be absorbed into an effective noise matrix:
\begin{equation}
    \Cmat = \Vmat_\mathrm{inf}^\top\, \Lmat_\mathrm{inf}\, \Vmat_\mathrm{inf} + \Neff,
    \label{eq:C_switched}
\end{equation}
where $\Vmat_\mathrm{inf}$ and $\Lmat_\mathrm{inf}$ contain only the modes with $\ell \in [\lmin, \lmax]$, and
\begin{equation}
    \Neff = \Nmat + \sum_{\substack{\ell \in [\lminsig, \lmin)\\\cup\, (\lmax, \lmaxsig]}} C_\ell^\mathrm{fid}\, \Wl\, \Pmat_\ell
    \label{eq:N_eff}
\end{equation}
is the instrumental noise augmented by the fixed signal contribution from both sides of the inference window. The SMW identity (Eq.~\ref{eq:smw}) is then applied with $\Neff$ in place of $\Nmat$, and the dimension of the kernel $\Kmat$ is reduced from $\nmodes(\lminsig, \lmaxsig)$ to $\nmodes(\lmin, \lmax)$. Since $\Neff$ is independent of the power spectrum being estimated, $\Neff^{-1}$ is computed only once. When switching is active, $\Nmat^{-1}$ itself is never assembled --- $\Neff^{-1}$ takes its place throughout Eqs.~\eqref{eq:smw}--\eqref{eq:weighted_data_smw} --- avoiding a redundant pixel-space inversion.

\subsection{Multipole binning}
\label{sec:binning}

On limited sky, the per-$\ell$ Fisher matrix can become singular or poorly conditioned and per-$\ell$ estimates carry strong inter-multipole correlations \citep{BondJaffeKnox1998, BilbaoAhedo2021}. \cf addresses both issues with native binned estimation adapted from \citet{Vanneste2018}: the bin-derivative $\Pmat_b = \sum_{\ell \in b} \Wl\, \Pmat_\ell$ replaces $\Pmat_\ell$ in the Fisher and $q$ formulae, and the algebra downstream is identical. The deconvolved bandpower $\hat{C}_b = (\Fmat^{-1}q)_b$ is then a Fisher-weighted combination of the $\Cl$ within the bin rather than a flat-$\Cl$ average, the two coinciding when $\Cl$ varies slowly across the bin. The implementation is a single code path: per-multipole estimation is the $\dbell = 1$ special case. Native binning is not equivalent to post-hoc averaging --- the two correspond to different estimators, $(\Pmat^\top \Fmat\, \Pmat)^{-1}\Pmat^\top \mathbf{q}$ vs $\Pmat\,\Fmat^{-1}\mathbf{q}$, which differ whenever $\Fmat$ has significant off-diagonal structure \citep{BondJaffeKnox1998}; the native version is preferred as it regularises the Fisher matrix and operates on the sufficient statistics directly.

\subsection{Normalisation modes}
\label{sec:normalisation}

The raw QML estimates $q_\ell$ (or $q_b$ in the binned case) can be presented in three different normalisations, each suited to different analysis needs. These correspond to the raw, deconvolved, and uncorrelated representations discussed in Sect.~V.A of \citet{Tegmark1997}. In the following we write the expressions for the per-$\ell$ case; the binned case is obtained by replacing $\ell$ indices with bin indices $b$.

\paragraph{Deconvolved mode.} The standard normalisation applies the inverse Fisher matrix to obtain estimates of the true power spectrum:
\begin{equation}
    \Chat^\mathrm{deconv} = \sum_{\ell'} (\Fmat^{-1})_{\ell\ell'}\, q_{\ell'}.
    \label{eq:deconvolved}
\end{equation}
The covariance of these estimates is $\mathrm{Cov}(\Chat, \hat{C}_{\ell'}) = (\Fmat^{-1})_{\ell\ell'}$, which is generally non-diagonal: estimates at nearby multipoles are correlated. This is the default output mode.

\paragraph{Decorrelated mode.} To obtain estimates with uncorrelated errors, the inverse square root of the Fisher matrix is applied \citep{Tegmark1997, TegmarkHamilton1997}:
\begin{equation}
    \Chat^\mathrm{decorr} = \sum_{\ell'} (\Fmat^{-1/2})_{\ell\ell'}\, q_{\ell'},
    \label{eq:decorrelated}
\end{equation}
where $\Fmat^{-1/2}$ is computed via eigendecomposition: $\Fmat = \Umat\Ximat\,\Umat^\top \Rightarrow \Fmat^{-1/2} = \Umat\,\Ximat^{-1/2}\,\Umat^\top$. Eigenvalues below $10^{-12}$ times the maximum are set to zero to regularise ill-conditioned modes. The covariance is the identity matrix by construction: $\mathrm{Cov}(\Chat^\mathrm{decorr}, \hat{C}_{\ell'}^\mathrm{decorr}) = \delta_{\ell\ell'}$, but the expectation value is $\langle\Chat^\mathrm{decorr}\rangle = \sum_{\ell'} (\Fmat^{1/2})_{\ell\ell'}\, C_{\ell'}$, not $C_\ell$ itself. That is, $\Fmat^{1/2}$ acts as a window function that mixes multipoles: the decorrelated estimates are uncorrelated linear combinations of the true power spectrum, not direct estimates of individual $C_\ell$. To compare with a theoretical model, one must apply the same $\Fmat^{1/2}$ transformation to the theory. These estimates are useful for visualisation (unit error bars at each data point) and for analyses that require independent measurements. The total information content is preserved --- it is merely redistributed across multipoles.

\paragraph{Convolved mode.} In this mode, the raw estimates are returned without any deconvolution, together with the window matrix $\Wmat_{\ell\ell'} = F_{\ell\ell'}$ that relates them to the true power spectrum:
\begin{equation}
    \langle q_\ell \rangle = \sum_{\ell'} \Wmat_{\ell\ell'}\, C_{\ell'}.
    \label{eq:convolved}
\end{equation}
Rather than inverting $\Wmat$ (which may be ill-conditioned), the user convolves the theoretical prediction: $C_\ell^\mathrm{th,conv} = \sum_{\ell'} \Wmat_{\ell\ell'}\, C_{\ell'}^\mathrm{th}$ and compares directly with $q_\ell$. The covariance of the raw estimates is $\Fmat$ itself. This mode is particularly useful when the Fisher matrix is poorly conditioned (e.g., for small sky fractions) or when the estimated spectra are to be used in a subsequent likelihood analysis that accounts for the window function explicitly.

\subsection{Pixel-space likelihood}
\label{sec:pixel_likelihood}

In addition to the QML power spectrum estimator, \cf implements a direct pixel-space Gaussian likelihood for cosmological parameter inference. Given a set of cosmological parameters $\params$ that determine the theoretical power spectra $C_\ell(\params)$, the log-likelihood of the data is
\begin{equation}
    \ln\mathcal{L}(\params) = -\frac{1}{2}\left[\dvec^\top \Cmat(\params)^{-1}\dvec + \ln|\Cmat(\params)|\right] + \mathrm{const},
    \label{eq:pixel_likelihood}
\end{equation}
where $\Cmat(\params) = \Smat(\params) + \Nmat$ and $\Smat(\params)$ is built from $C_\ell(\params)$ via Eq.~\eqref{eq:signal_cov}. Both the quadratic form and the log-determinant can be evaluated efficiently using the SMW identity (Eqs.~\ref{eq:smw} and~\ref{eq:logdet}), reducing the per-point cost from $\mathcal{O}(\npix^3)$ to $\mathcal{O}(\nmodes^3)$. When $\ell$-switching is employed (Sect.~\ref{sec:ell_switching}), the cost is further reduced to $\mathcal{O}(\nmodes(\lmin, \lmax)^3)$.

In \cf, the likelihood is evaluated over a Cartesian grid of cosmological parameters. At each grid point, the theoretical power spectra are loaded from precomputed files (e.g., generated by {\tt CAMB} or {\tt CLASS}), the signal covariance is rebuilt, and the likelihood is evaluated. The grid points are distributed across MPI processes in a round-robin fashion for near-perfect load balancing. Results are collected into a {\tt LikelihoodResult} object that provides best-fit parameter extraction, marginalised likelihoods, and confidence interval computation.

The QML estimator (Eq.~\ref{eq:qml_estimate}) corresponds to a single Newton--Raphson step towards the maximum of Eq.~\eqref{eq:pixel_likelihood}, starting from the fiducial model \citep{BondJaffeKnox1998}: the pixel-space likelihood is therefore the exact target that the single-step QML estimator approximates and to which iterative QML converges.
\section{Code architecture and implementation}
\label{sec:architecture}

\cf is implemented in Python and organised as three interconnected packages with a layered architecture: \cc provides the computational foundation, \qube implements the QML estimation pipeline, and \pics evaluates the pixel-space likelihood. Each package is independently installable but shares a common infrastructure through \cc. In this section we describe the design principles, the main classes and their responsibilities, and the computational strategies employed.

\subsection{Package structure}
\label{sec:packages}

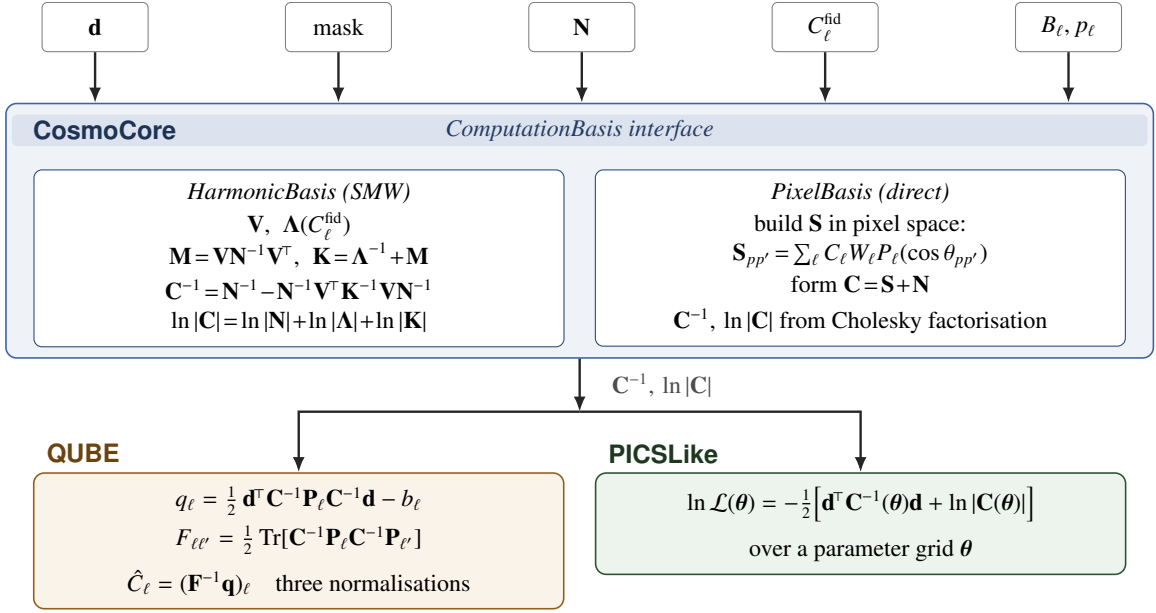
\begin{figure*}
    \centering
%
%
\begin{tikzpicture}[
    >={Latex[length=2mm,width=1.6mm]},
    font=\small,
    input/.style={
        draw=black!50, fill=white, line width=0.4pt,
        rounded corners=2pt, font=\small,
        minimum width=14mm, minimum height=6.5mm, align=center
    },
    basisbox/.style={
        draw=CFblue!70!black, fill=white, line width=0.4pt,
        rounded corners=3pt, align=center,
        inner xsep=5pt, inner ysep=4pt, font=\footnotesize,
        minimum width=70mm
    },
    qubeout/.style={
        draw=CForange!70!black, fill=CForange!10, line width=0.6pt,
        rounded corners=4pt, font=\footnotesize, align=center,
        inner xsep=6pt, inner ysep=5pt, minimum width=70mm
    },
    picsout/.style={
        draw=CFgreen!60!black, fill=CFgreen!10, line width=0.6pt,
        rounded corners=4pt, font=\footnotesize, align=center,
        inner xsep=6pt, inner ysep=5pt, minimum width=70mm
    },
    pkglabel/.style={
        font=\bfseries\sffamily, inner xsep=3pt, inner ysep=1pt
    },
    busline/.style={line width=0.5pt, black!65},
    flowmain/.style={->, line width=0.9pt, black!85}
]

\node[font=\footnotesize\itshape, CFblue!60!black,
      fill=CFblue!18, rounded corners=2pt, inner xsep=5pt, inner ysep=2pt,
      minimum width=150mm, align=center]
     (subtitle) at (0, 0) {ComputationBasis interface};

\node[basisbox, anchor=north east] (harm)
     at ([xshift=-2mm, yshift=-3mm]subtitle.south) {%
       \begin{tabular}{c}
       \itshape HarmonicBasis (SMW)
       \\[2pt]
       $\mathbf{V}$,\;
       $\boldsymbol{\Lambda}(C_\ell^\mathrm{fid})$
       \\[1pt]
       $\mathbf{M}\!=\!\mathbf{V}\mathbf{N}^{-1}\mathbf{V}^{\!\top}$,\;
       $\mathbf{K}\!=\!\boldsymbol{\Lambda}^{-1}\!+\!\mathbf{M}$
       \\[3pt]
       $\mathbf{C}^{-1}\!=\!\mathbf{N}^{-1}\!-\!\mathbf{N}^{-1}\mathbf{V}^{\!\top}\mathbf{K}^{-1}\mathbf{V}\mathbf{N}^{-1}$
       \\[1pt]
       $\ln|\mathbf{C}|\!=\!\ln|\mathbf{N}|\!+\!\ln|\boldsymbol{\Lambda}|\!+\!\ln|\mathbf{K}|$
       \end{tabular}};

\node[basisbox, anchor=north west] (pix)
     at ([xshift=2mm, yshift=-3mm]subtitle.south) {%
       \begin{tabular}{c}
       \itshape PixelBasis (direct)
       \\[2pt]
       build $\mathbf{S}$ in pixel space:
       \\[1pt]
       $\mathbf{S}_{pp'}\!=\!\sum_\ell C_\ell W_\ell P_\ell(\cos\theta_{pp'})$
       \\[2pt]
       form $\mathbf{C}\!=\!\mathbf{S}\!+\!\mathbf{N}$
       \\[3pt]
       $\mathbf{C}^{-1},\,\ln|\mathbf{C}|$ from Cholesky factorisation
       \end{tabular}};

\begin{scope}[on background layer]
    \node[
        fit=(subtitle)(harm)(pix),
        inner xsep=2pt, inner ysep=4pt,
        draw=CFblue, line width=0.6pt,
        fill=CFblue!8, rounded corners=4pt
    ] (cosmocore_box) {};
\end{scope}

\node[pkglabel, CFblue!50!black, anchor=west]
     at ([xshift=5pt]subtitle.west) {CosmoCore};

\node[input] (data) at ([xshift=-64mm, yshift=10mm]cosmocore_box.north)
     {$\mathbf{d}$};
\node[input, right=18mm of data] (mask) {mask};
\node[input, right=18mm of mask] (N)    {$\mathbf{N}$};
\node[input, right=18mm of N]    (Cl)   {$C_\ell^\mathrm{fid}$};
\node[input, right=18mm of Cl]   (Bl)   {$B_\ell, p_\ell$};

\foreach \n in {data, mask, N, Cl, Bl} {
    \draw[flowmain] (\n.south) -- (\n.south |- cosmocore_box.north);
}

\node[qubeout, anchor=north east] (qube_out)
     at ([xshift=-2mm, yshift=-15mm]cosmocore_box.south)
     {%
       $q_\ell = \tfrac{1}{2}\,\mathbf{d}^{\!\top}\mathbf{C}^{-1}\mathbf{P}_\ell\mathbf{C}^{-1}\mathbf{d} - b_\ell$
       \\[3pt]
       $F_{\ell\ell'} = \tfrac{1}{2}\,\mathrm{Tr}[\mathbf{C}^{-1}\mathbf{P}_\ell \mathbf{C}^{-1}\mathbf{P}_{\ell'}]$
       \\[5pt]
       $\hat{C}_\ell = (\mathbf{F}^{-1}\mathbf{q})_\ell$\quad
       \footnotesize three normalisations
     };

\node[picsout, anchor=north west] (pics_out)
     at ([xshift=2mm, yshift=-15mm]cosmocore_box.south)
     {%
       $\ln\mathcal{L}(\boldsymbol{\theta}) = -\tfrac{1}{2}\!\left[\mathbf{d}^{\!\top}\mathbf{C}^{-1}(\boldsymbol{\theta})\mathbf{d} + \ln|\mathbf{C}(\boldsymbol{\theta})|\right]$
       \\[6pt]
       \footnotesize over a parameter grid $\boldsymbol{\theta}$
     };

\node[pkglabel, CForange!50!black, anchor=south west]
     at ([xshift=2pt, yshift=2pt]qube_out.north west) {QUBE};
\node[pkglabel, CFgreen!40!black, anchor=south west]
     at ([xshift=2pt, yshift=2pt]pics_out.north west) {PICSLike};

\coordinate (fork) at ([yshift=-7mm]cosmocore_box.south);
\draw[flowmain] (cosmocore_box.south) -- (fork);
\draw[flowmain] (fork) -| (qube_out.north);
\draw[flowmain] (fork) -| (pics_out.north);

\node[font=\footnotesize\itshape, black!70, anchor=west]
     at ([xshift=3mm, yshift=-3.5mm]cosmocore_box.south)
     {$\mathbf{C}^{-1},\,\ln|\mathbf{C}|$};

\end{tikzpicture}
    \caption{Architecture and data flow of \cf. The standard inputs (top: data $\mathbf{d}$, mask, noise covariance $\mathbf{N}$, fiducial power spectrum $C_\ell^\mathrm{fid}$, beam and pixel windows $B_\ell, p_\ell$) feed \cc, which exposes a common {\tt ComputationBasis} interface with two interchangeable implementations --- the harmonic basis (left), where $\mathbf{C}^{-1}$ and $\ln|\mathbf{C}|$ are obtained from the Sherman--Morrison--Woodbury identity and the matrix determinant lemma applied to the operators $\mathbf{V}, \boldsymbol{\Lambda}, \mathbf{M}, \mathbf{K}$ (Sect.~\ref{sec:harmonic}); and the pixel basis (right), where $\mathbf{S}$ is built directly in pixel space, $\mathbf{C}\!=\!\mathbf{S}\!+\!\mathbf{N}$ is formed explicitly, and $\mathbf{C}^{-1}$ together with $\ln|\mathbf{C}|$ are obtained from a Cholesky factorisation. Both paths drive QML power spectrum estimation in \qube (left) and pixel-space likelihood evaluation in \pics (right).}
    \label{fig:architecture}
\end{figure*}

The three packages are:

\begin{itemize}
    \item \cc --- Core computational library. Provides field management (spin-0 and spin-2), spherical harmonic and pixel-space operations, matrix algebra, the harmonic operator $\Vmat$ and SMW machinery, binning, beam handling, and I/O utilities. All numerically intensive operations (Legendre polynomial evaluation, Wigner $d$-matrix computation, signal matrix construction) are accelerated with Numba JIT compilation \citep{numba} to achieve performance comparable to compiled languages.

    \item \qube (Quadratic maximum likelihood UnBiased Estimator) --- QML analysis engine. Implements the {\tt Fisher} and {\tt Spectra} classes for Fisher matrix computation and power spectrum estimation, including all three normalisation modes (Sect.~\ref{sec:normalisation}), noise bias computation, and cross-spectrum support.

    \item \pics (Pixel-based Inference with Correlated-Skies Likelihood) --- Pixel-space likelihood engine. Implements the {\tt PICSLike} class for evaluating the Gaussian likelihood (Eq.~\ref{eq:pixel_likelihood}) over parameter grids, with automated marginalisation and confidence interval extraction.
\end{itemize}

Both \qube and \pics inherit from the {\tt Core} abstract base class defined in \cc, which provides the common setup pipeline: field initialisation, pixel geometry computation, noise covariance loading, power spectrum and beam configuration, and --- when the harmonic-space formulation is used --- construction of the $\Vmat$ operator and SMW quantities. This shared inheritance ensures that the same covariance infrastructure is used consistently across QML estimation and likelihood evaluation. A schematic overview of the package structure is shown in Fig.~\ref{fig:architecture}.

\subsection{CosmoCore}
\label{sec:cosmocore}
\label{sec:harmonic_basis_code}

\cc exposes a type system of field classes ({\tt ScalarField} spin-0, {\tt PolarizationField} spin-2, both inheriting from {\tt BaseField}) and a {\tt FieldCollection} that groups them into a joint analysis, auto-derives the relevant auto- and cross-spectra, and lets each component carry its own multipole floor (so the spin-0 monopole or dipole can be estimated jointly with $\ell\!\ge\!2$ polarisation in a single TQU pass; see Appendix~\ref{app:yaml}). Low-level mathematics --- Legendre polynomials with absorbed $(2\ell+1)/(4\pi)$ normalisation, Wigner-$d$ matrices via Jacobi recurrence (Appendix~\ref{app:spin_harmonics}), Cholesky-based linear algebra, and the SMW identity (Appendix~\ref{app:smw_logdet}) --- lives in the {\tt basics} subpackage and is JIT-compiled with Numba. The pixel-space signal covariance $\Smat$ and its per-$\ell$ derivatives are built by the {\tt pixel} module with separate spin-0$\times$0, spin-2$\times$2 and spin-0$\times$2 paths.

\paragraph{Computation basis.} The harmonic-vs-pixel choice is exposed through a {\tt ComputationBasis} abstract interface with two implementations: {\tt HarmonicBasis} carries $\Vmat$, $\Lmat$, $\Mmat\!=\!\Vmat\Nmat^{-1}\Vmat^\top$ and $\Kmat$, and reduces $\Ctinv$, the Fisher matrix, and the likelihood terms to operations on those objects via the SMW identity; {\tt PixelBasis} forms $\Cmat$ explicitly and Cholesky-inverts it. The \texttt{auto} selector compares the leading-order costs --- $\nmodes^3$ for the SMW kernel against $(\nbins+1)\,\npix^3$ for the per-bin $\Cmat^{-1}\,\partial\Cmat/\partial\Cl$ products --- and picks the cheaper basis at setup time. When $\ell$-switching is active (Sect.~\ref{sec:ell_switching}), the effective $\nmodes$ shrinks and the harmonic path remains applicable to higher resolutions.

\subsection{QUBE}
\label{sec:qube_code}

The \qube package implements the QML estimation pipeline through two main classes.

The {\tt Fisher} class computes the Fisher information matrix (Eq.~\ref{eq:fisher} or~\ref{eq:fisher_harmonic}) and optionally caches the binned derivative matrices for reuse by the {\tt Spectra} class. It supports both single-spectrum (e.g., a scalar-only analysis) and multi-spectrum (e.g., joint scalar and spin-2) configurations. In the multi-spectrum case, the Fisher matrix has dimension $(\nspec \times \nbins)^2$, with blocks coupling different spectra and multipole bins. The computation is parallelised across MPI processes by distributing the Fisher matrix element pairs.

The {\tt Spectra} class computes the QML power spectrum estimates $q_\ell$ (Eq.~\ref{eq:q_ell} or~\ref{eq:q_harmonic}), the noise bias $b_\ell$ (Eq.~\ref{eq:noise_bias}), and returns the final estimates in any of the three normalisation modes (Sect.~\ref{sec:normalisation}). It can accept a precomputed {\tt Fisher} instance to avoid redundant computation, and inherits the cached derivatives and basis configuration. Multiple simulation maps are processed together at every rank, vectorised across the simulation index through {\tt numpy} broadcasting; the MPI parallelism on this stage distributes the (spectrum, bin) pairs of the QML evaluation across ranks rather than the simulations themselves. The cross-correlation mode (Eq.~\ref{eq:q_cross}) is supported for pairs of independent data sets with uncorrelated noise.

\subsection{PICSLike}
\label{sec:picslike_code}

The \pics package evaluates the pixel-space Gaussian likelihood (Eq.~\ref{eq:pixel_likelihood}) across a user-specified grid of cosmological parameters.

The {\tt ParameterGrid} class manages the Cartesian product of parameter ranges and the associated theoretical power spectra, which are loaded from precomputed files at each grid point. The $\ell$-switching mode (Sect.~\ref{sec:ell_switching}), shared with the QML pipeline through the common {\tt Core} setup, lets the user vary the power spectrum only over a target multipole range while holding higher multipoles fixed at a fiducial model. This reduces the multipole range that must be tabulated in each grid-point spectra file and the per-point computational cost; the file count is unchanged.

At each grid point, the signal covariance $\Smat(\params)$ is rebuilt from the theoretical $\Cl(\params)$, the total covariance $\Cmat(\params) = \Smat(\params) + \Nmat$ is formed, and the log-likelihood is evaluated. When the harmonic basis is available, the SMW identity provides both the quadratic form $\dvec^\top\Cmat^{-1}\dvec$ and the log-determinant $\ln|\Cmat|$ at a cost of $\mathcal{O}(\nmodes^3)$ per grid point. Grid points are distributed across MPI processes in a round-robin fashion. Multiple simulation maps are evaluated together at every rank, sharing the per-point covariance inverse and producing per-simulation likelihood curves at no extra inversion cost.

The {\tt LikelihoodResult} class stores the $\chi^2$ and log-likelihood values across the parameter grid, and provides methods for best-fit extraction, marginalised likelihoods (by summing over unwanted parameter dimensions), and confidence interval computation (by finding the smallest interval containing the desired probability mass). Results are serialisable for later analysis.

The configuration interface that drives every \cf analysis (YAML schema, default templates, and a minimal Python usage example) is documented in Appendix~\ref{app:yaml}.

\section{Validation}
\label{sec:validation}

In this section we present the validation of \cf against the Planck low-$\ell$ Fortran reference implementation and through internal consistency tests. The validation covers every stage of the pipeline: signal covariance construction, covariance inversion, Fisher matrix computation, QML power spectrum estimation, noise bias, and pixel-space likelihood evaluation. All tests use HEALPix maps at $\nside = 4$--$8$ with realistic masking and noise configurations.

\subsection{Planck low-$\ell$ reference validation}
\label{sec:fortran_validation}

We validate \cf against the Planck low-$\ell$ Fortran reference of \citet{Pagano2020} and the {\tt simall} likelihood of \citet{PlanckV2020}\footnote{The {\tt pse\_qml} (QML estimation) and {\tt parameter\_estimation} (pixel-space likelihood) codes are at \url{https://baltig.infn.it/cosmology_ferrara/lowell-likelihood-analysis}.} element-by-element on a $B$-mode analysis at $\nside = 8$, $\lmax = 24$, with a tabulated beam and white noise. The primary covariance products ($\Nmat$, $\Smat$, $\Cmat^{-1}$) reproduce the reference at machine precision (10$^{-11}$--10$^{-13}$, $\Nmat$ bit-exact); the trace-based final products of the QML pipeline ($\Fmat$, $\Chat$, $b_\ell$) sit at $\sim\!10^{-7}$ and those of the pixel-space likelihood pipeline ($\chi^2$, $\ln\mathcal{L}$) at $\sim\!10^{-9}$, the latter tighter because it carries fewer sequential trace evaluations; the floors reflect floating-point accumulation order between the Numba/NumPy and Fortran summations rather than any algorithmic disagreement. The same picture is recovered in spin-2: against {\tt dofisher\_QU} + {\tt qml\_QU} at $\nside = 8$, $\lmax = 32$ with a Gaussian beam, the noise-bias rel-diff stays $\le 6\times 10^{-6}$ for $EE$ ($1\times 10^{-9}$ for $BB$), with median $\sim 10^{-10}$.

The harmonic and pixel-space computation paths agree internally to $\sim\!10^{-13}$ on the Fisher matrix in both single-field spin-0 and multi-field configurations, confirming that the SMW identity and the spin-2 machinery (Wigner-$d$, $2{\times}2$ $\Lmat$ blocks, rotation angles) are correctly assembled. Structural sanity checks of the spin-2 path --- $\Vmat$ block shape ($2\nmodes^\mathrm{base} \times 2\npix^\mathrm{phys}$ with non-zero $E$ and $B$ sub-blocks), $2{\times}2$ $\Lmat$ blocks with the correct $EE/BB/EB$ entries, and a multi-field benchmark on three independent spin-0 fields simulating $T,E,B$ --- hold at the same precision; the native-binning path is validated by $\Fmat^\mathrm{binned} = \Pmat_\Sigma \Fmat^\mathrm{unbinned} \Pmat_\Sigma^\top$ and $q_b = \sum_{\ell \in b} q_\ell$ to machine precision for $\dbell = 2, 3, 5$.

\subsection{Monte Carlo validation}
\label{sec:monte_carlo}

We validate the statistical properties of the QML estimator on $\Nsim = 10\,000$ Gaussian polarization $(Q, U)$ realisations at $\nside = 16$, $\ell \in [2, 32]$, drawn from a $\Lambda$CDM lensing-only model ($\tau = 0.06$, $r = 0$) with a cosine-window beam \citep{Benabed2009} applied at the harmonic level and white pixel noise at $\sigma = 10^{-3}\,\mu\mathrm{K\,arcmin}$ matching the QML noise covariance $\Nmat = \sigma^2 \Imat$. A $\pm 10^\circ$ Galactic cut yields $\fsky \approx 89\%$, and the analysis estimates $EE$, $BB$, $EB$ jointly in the harmonic basis with noise-bias subtraction.

\begin{figure}
    \centering
    \includegraphics[width=\columnwidth]{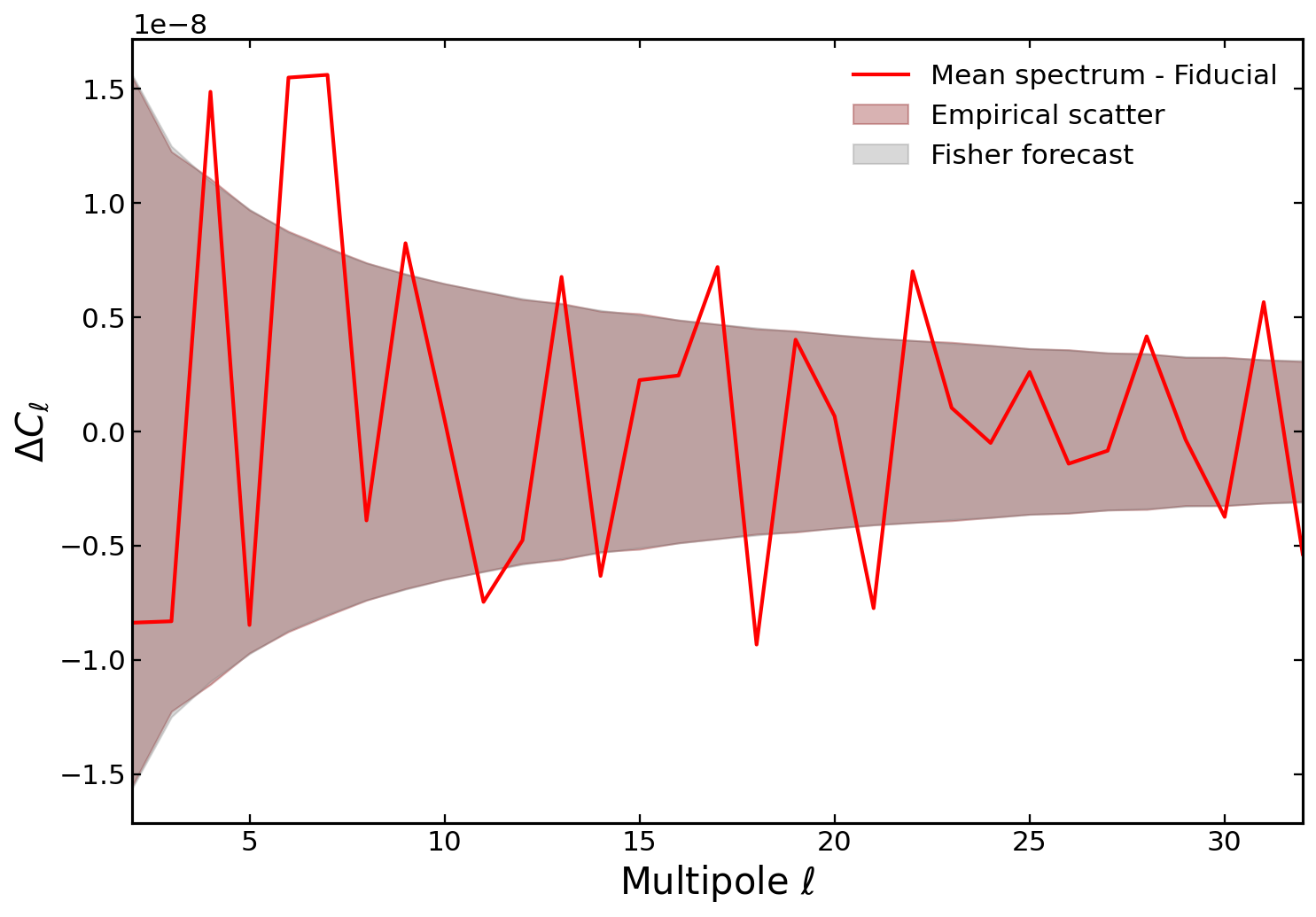}
    \caption{Monte Carlo validation of the QML estimator for the $BB$ power spectrum. The red line shows the difference between the mean recovered spectrum and the input fiducial over $10\,000$ realisations. The red band is the standard error on the mean ($\sigma_\ell / \sqrt{\Nsim}$), and the grey band is the Fisher-predicted uncertainty scaled by $1/\sqrt{\Nsim}$. The two bands overlap, confirming that the Fisher matrix correctly predicts the estimator variance. Configuration: $\nside = 16$, $\lmax = 32$, spin-2 ($Q$, $U$) analysis, cosine window beam, $\fsky \approx 89\%$.}
    \label{fig:unbiasedness}
\end{figure}

\paragraph{Unbiasedness.} Figure~\ref{fig:unbiasedness} shows the difference between the mean recovered $BB$ spectrum over the $10\,000$ realisations and the input fiducial, compared to the standard error on the mean $\sigma_\ell / \sqrt{\Nsim}$ (red band) and the Fisher-predicted uncertainty $\sqrt{(\Fmat^{-1})_{\ell\ell}} / \sqrt{\Nsim}$ (grey band). The mean residual is consistent with zero across all multipoles, with no systematic trend, confirming that the QML estimator is unbiased at the $\sim 10^{-8}$ precision level. The $BB$ spectrum is the most demanding test because it is subdominant to $EE$ by several orders of magnitude, and any E-to-B leakage through the mask would appear as a positive bias in $BB$. The absence of such leakage confirms that the full pixel-space covariance treatment handles E/B separation correctly.

\begin{figure}
    \centering
    \includegraphics[width=\columnwidth]{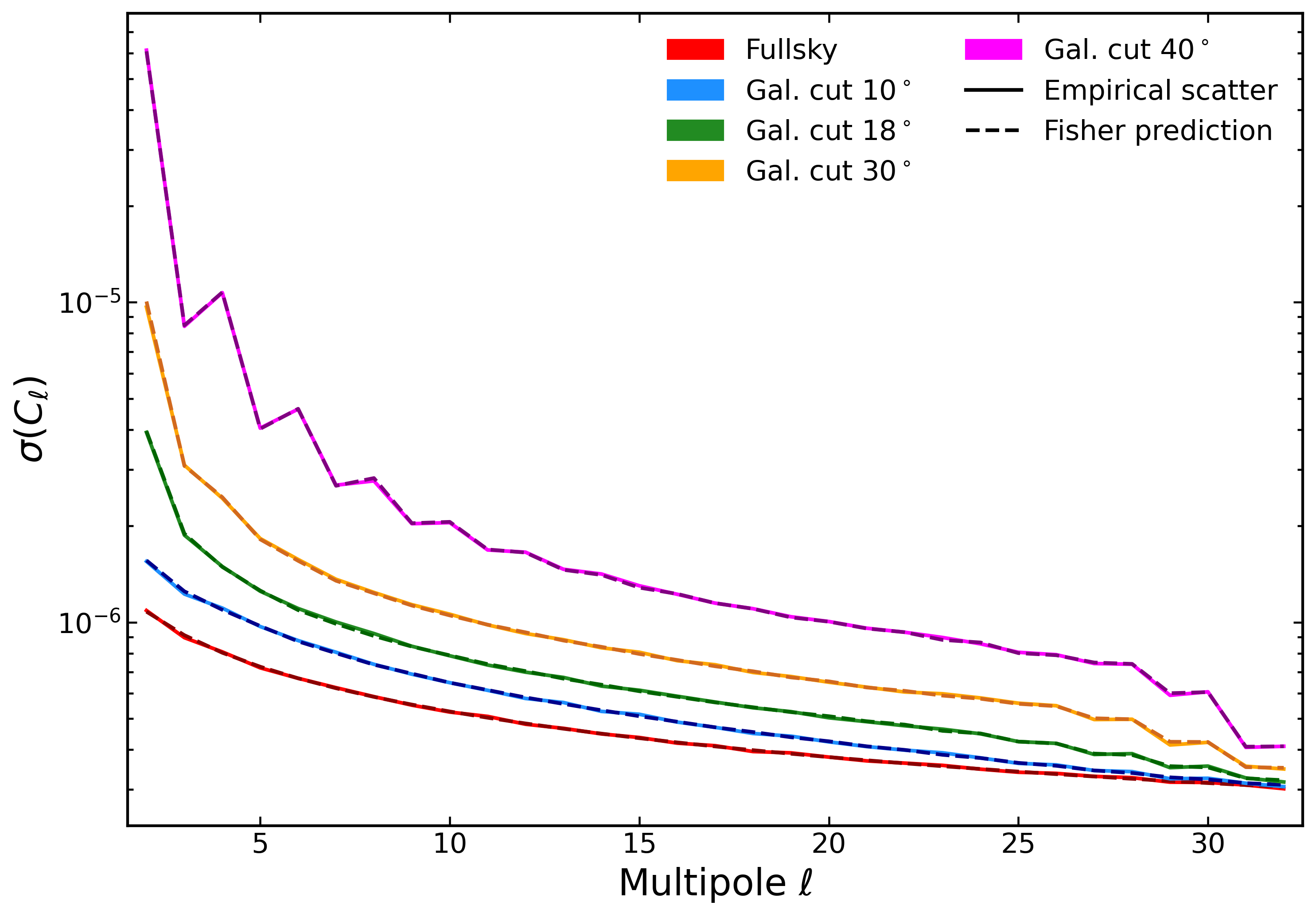}
    \caption{Standard deviation of the estimated $BB$ power spectrum over $10\,000$ MC realisations (solid lines) compared to the Fisher-predicted error bars (dashed lines), for five sky fractions: full sky (red), $\pm 10^\circ$ Galactic cut (blue), $\pm 18^\circ$ (green), $\pm 30^\circ$ (orange), and $\pm 40^\circ$ (magenta). Configuration: $\nside = 16$, $\lmax = 32$, spin-2 ($Q$, $U$) analysis.}
    \label{fig:multi_fsky}
\end{figure}

\paragraph{Variance and sky fraction dependence.} Figure~\ref{fig:multi_fsky} compares the empirical standard deviation of the $BB$ power spectrum estimates against the Fisher-predicted error bars $\sqrt{(\Fmat^{-1})_{\ell\ell}}$ for five sky fractions, from full sky to a $\pm 40^\circ$ Galactic cut ($\fsky \approx 36\%$). Empirical scatter and Fisher prediction overlap at every multipole and every $\fsky$, confirming that $\Fmat^{-1}$ accurately captures the estimator variance regardless of mask geometry; the combined window function $\Wl = \Bl^2 \pl^2$ is absorbed into the Fisher derivatives so that no post-hoc beam correction is required.

\paragraph{Noise bias subtraction.} The simulations include small white noise ($\sigma = 10^{-3}\,\mu\mathrm{K\,arcmin}$). At this level the noise is well below the cosmic-variance amplitude of the recovered spectra at every multipole considered; its role here is regularisation, providing finite eigenvalues for the modes of $\Cmat$ that the beam window function $\Wl$ would otherwise drive towards zero in the signal-only limit and which would render $\Cmat$ numerically singular. The same $\Nmat$ is used both to draw the noise realisations and to construct the analytical noise bias $b_\ell = \frac{1}{2}\,\mathrm{Tr}[\Cmat^{-1} \Emat_\ell \Cmat^{-1} \Nmat]$ (Eq.~\ref{eq:noise_bias}), and the unbiasedness reported in Fig.~\ref{fig:unbiasedness} therefore validates the noise bias computation and its subtraction simultaneously. A complementary noise-dominated Monte Carlo run --- the polarisation $BB$ analogue of the temperature-noise regime partially sampled by the QML-vs-PCL comparison of Sect.~\ref{sec:qml_vs_pcl}, where the noise drives the variance rather than acting as a regulariser --- is in preparation and will be presented in a dedicated companion note.

\subsection{End-to-end pixel-based inference}
\label{sec:picslike_inference}

To close the validation loop and make the unification claim of Sect.~\ref{sec:formalism} concrete, we drive the \pics likelihood end-to-end on the very same Monte Carlo sample used in Fig.~\ref{fig:unbiasedness}. The simulations, the noise covariance, the mask, the beam, and the resolution are reused without modification ($\nside = 16$, $\lmax = 32$, cosine-window beam, $\fsky \approx 0.89$, $r_\mathrm{true} = 0$); only the analysis target changes, from per-multipole bandpowers to a 1D scan of the tensor-to-scalar ratio $r$ over the physical range $r \in [0,\, 3 \times 10^{-4}]$ (31 grid points). The grid spectra are built from the same fiducial $\Lambda$CDM lensing baseline that generated the simulations, with a linear tensor BB template on top.

\begin{figure}
    \centering
    \includegraphics[width=\columnwidth]{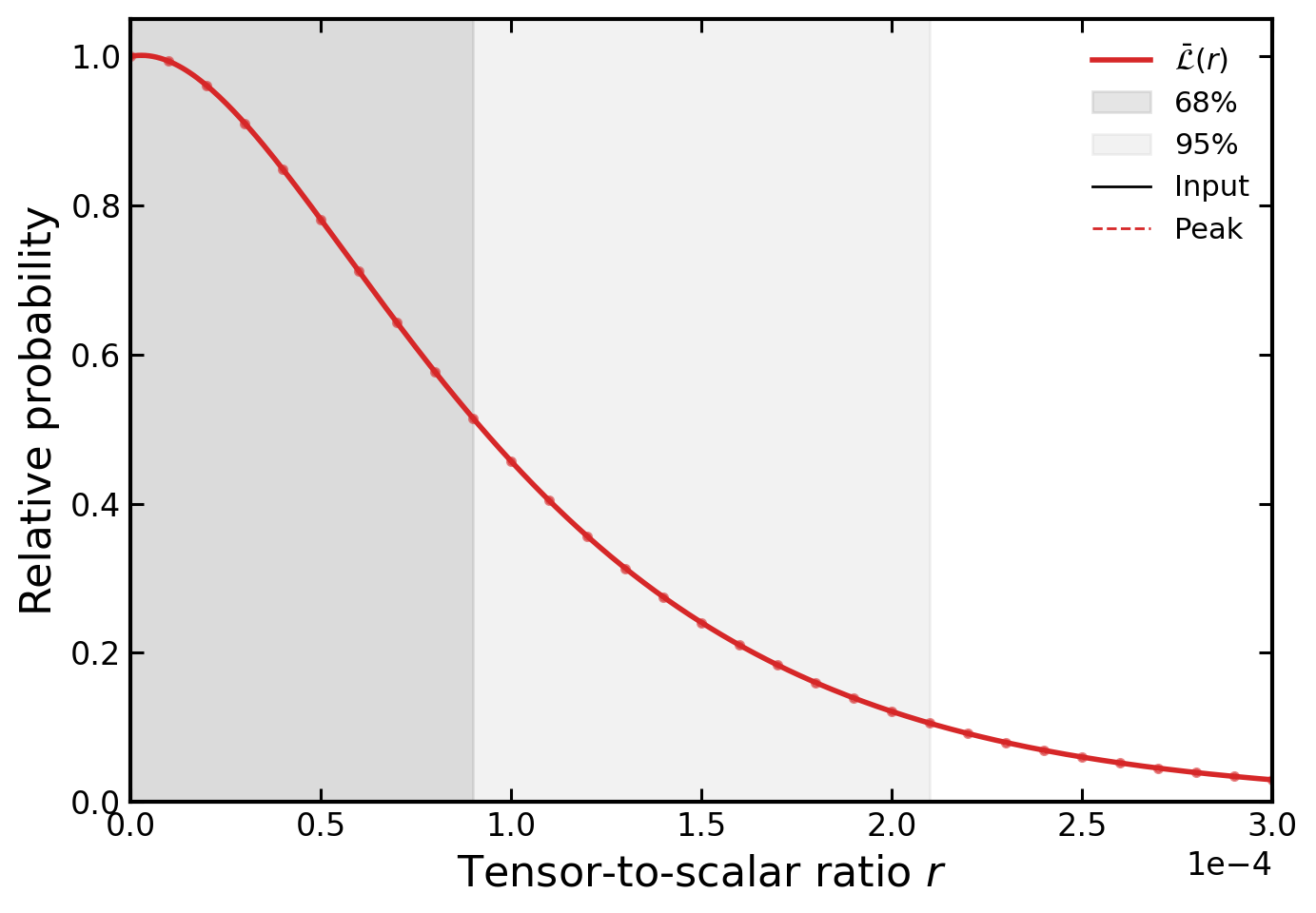}
    \caption{Mean-$\chi^2$ \pics likelihood $\bar{\mathcal{L}}(r) \propto \exp[-\tfrac{1}{2}\langle\chi^2(r)\rangle_\mathrm{sims}]$ on the tensor-to-scalar ratio $r$ from $\Nsim = 1000$ Monte Carlo simulations, evaluated on the same configuration as Fig.~\ref{fig:unbiasedness} ($\nside = 16$, $\lmax = 32$, cosine-window beam, $\fsky \approx 0.89$, input $r_\mathrm{true} = 0$). Markers indicate the $31$ \pics grid points; the solid curve is a cubic-spline interpolation between them. The peak sits at the lower edge of the physical range $r \geq 0$, with $68\%$ and $95\%$ upper limits $r < 9 \times 10^{-5}$ and $r < 2.1 \times 10^{-4}$ (shaded bands), consistent with the input within the MC sampling error.}
    \label{fig:picslike_r_posterior}
\end{figure}

Figure~\ref{fig:picslike_r_posterior} shows the mean-$\chi^2$ likelihood
\begin{equation}
    \bar{\mathcal{L}}(r)
    \;\propto\;
    \exp\!\left[-\tfrac{1}{2}\,\langle \chi^2(r) \rangle_\mathrm{sims}\right],
    \label{eq:mean_chi2_likelihood}
\end{equation}
constructed from the per-realisation $\chi^2(r)$ across $\Nsim = 1000$ Monte Carlo simulations. We stress that $\bar{\mathcal{L}}(r)$ is \emph{not} the sample average of the per-realisation likelihoods $\langle \mathcal{L}(r) \rangle_\mathrm{sims}$: the two coincide only in the limit of small $\chi^2$ fluctuations across realisations, since by Jensen's inequality \citep{Jensen1906} $\langle \exp(-\chi^2/2) \rangle \ge \exp(-\langle \chi^2 \rangle/2)$. We use $\bar{\mathcal{L}}$ here because, for an unbiased estimator with a correct covariance model, $\mathbb{E}[\chi^2(r)] = \mathrm{Tr}[\Cmat(r)^{-1}\,\Cmat_\mathrm{true}] + \ln|\Cmat(r)|$ is minimised exactly at $r = r_\mathrm{true}$, so the peak of Eq.~\eqref{eq:mean_chi2_likelihood} is itself the unbiasedness test. With the prior $r \geq 0$, the recovered posterior peaks at the lower edge of the grid, $r = 0 = r_\mathrm{true}$, demonstrating unbiasedness, and yields one-sided upper limits $r < 9 \times 10^{-5}$ at $68\%$ credibility and $r < 2.1 \times 10^{-4}$ at $95\%$.

No additional infrastructure is required beyond the QML pipeline: the pixel-space covariance $\Cmat = \Smat(r) + \Nmat$ that supplies the Fisher matrix and noise bias for \qube\ also supplies the quadratic form and log-determinant of the pixel-space likelihood, with the same SMW machinery (Sect.~\ref{sec:harmonic}, Appendix~\ref{app:smw_logdet}) accelerating both. Application to real CMB data and a comparison against external pixel-space likelihood codes is the subject of a forthcoming companion paper.

\subsection{Optimality versus pseudo-$C_\ell$}
\label{sec:qml_vs_pcl}

To benchmark the optimality gap that motivates the QML approach, we compare \cf\ against the public {\tt NaMaster} pseudo-$C_\ell$ implementation \citep{Alonso2019} on the same set of simulations. The test is in temperature ($TT$) at $\nside = 32$, $\lmax = 2\nside = 64$, with $\Nsim = 1000$ Gaussian CMB+noise realisations. The white-noise level is set to $\sigma = 1.5\,\mu\mathrm{K}$ per pixel ($\sim 520\,\mu\mathrm{K\,arcmin}$ in $TT$ angular sensitivity), chosen to match the per-mode S/N of a $2\,\mu\mathrm{K\,arcmin}$ polarisation experiment on $BB$ at $\ell = 50$, so the temperature analysis sits in the same regime that a polarisation experiment faces on its primary $B$-mode target. Two Galactic-strip masks bracket the relevant range: a $24^\circ$ cut yielding $\fsky \approx 0.60$ analysed at $\Delta\ell = 1$, and a $64^\circ$ cut yielding $\fsky \approx 0.10$ analysed at $\Delta\ell = 5$. The $\lmax = 2\nside$ choice keeps the analysis inside the HEALPix-exact regime where the pseudo-$C_\ell$ transform is unbiased, ensuring the comparison isolates the optimality gap rather than transform inaccuracy.

\begin{figure*}
    \centering
    \includegraphics[width=\textwidth]{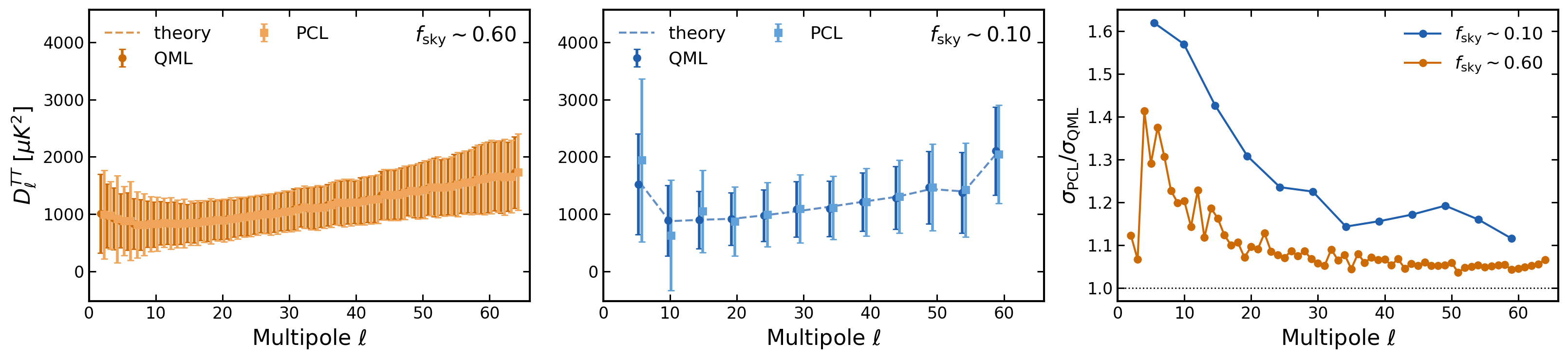}
    \caption{Comparison of \cf's QML estimator and the public {\tt NaMaster} pseudo-$C_\ell$ implementation \citep{Alonso2019} over $\Nsim = 1000$ simulations of CMB-plus-noise $TT$ maps at $\nside = 32$, $\lmax = 64$, $\sigma = 1.5\,\mu\mathrm{K}$ per pixel ($\sim 520\,\mu\mathrm{K\,arcmin}$ in $TT$, $TT$-equivalent of $2\,\mu\mathrm{K\,arcmin}$ polarisation noise at $\ell=50$), for two Galactic-strip masks ($\fsky \approx 0.10$ and $0.60$). \emph{Left:} mean recovered $D_\ell^{TT}$ bandpowers at $\fsky \approx 0.60$ (per-multipole estimation, $\dbell=1$) compared to the windowed theory; error bars are the empirical per-realisation standard deviation. \emph{Middle:} same at $\fsky \approx 0.10$ (binned estimation, $\dbell=5$). \emph{Right:} per-bandpower ratio $\sigma_\mathrm{PCL}/\sigma_\mathrm{QML}$ for both cases; the QML estimator is everywhere tighter, with the gap reaching $\sim 60\%$ at the lowest multipoles of the small-$\fsky$ case.}
    \label{fig:qml_vs_pcl_variance}
\end{figure*}

\paragraph{Recovery and variance.} Figure~\ref{fig:qml_vs_pcl_variance} (left) shows the recovered $D_\ell^{TT}$ bandpowers; both methods agree with the windowed theory across both sky fractions. The right panel reports the per-bandpower ratio $\sigma_\mathrm{PCL}/\sigma_\mathrm{QML}$. As expected, at large scales the QML estimator strongly outperforms pseudo-$C_\ell$: at $\fsky \approx 0.10$ the pseudo-$C_\ell$ error bars are $\sim 35$--$60\%$ larger than the QML ones across the lowest six bandpowers ($\ell \lesssim 30$), and even at $\fsky \approx 0.60$ the pseudo-$C_\ell$ variance exceeds the optimal one by up to $\sim 40\%$ at the lowest multipoles. At smaller scales ($\ell \gtrsim 30$) the gap shrinks but does not close: pseudo-$C_\ell$ is still $\sim 15$--$20\%$ noisier than QML at $\fsky \approx 0.10$, and a few percent noisier at $\fsky \approx 0.60$.

\begin{figure*}
    \centering
    \includegraphics[width=0.65\textwidth]{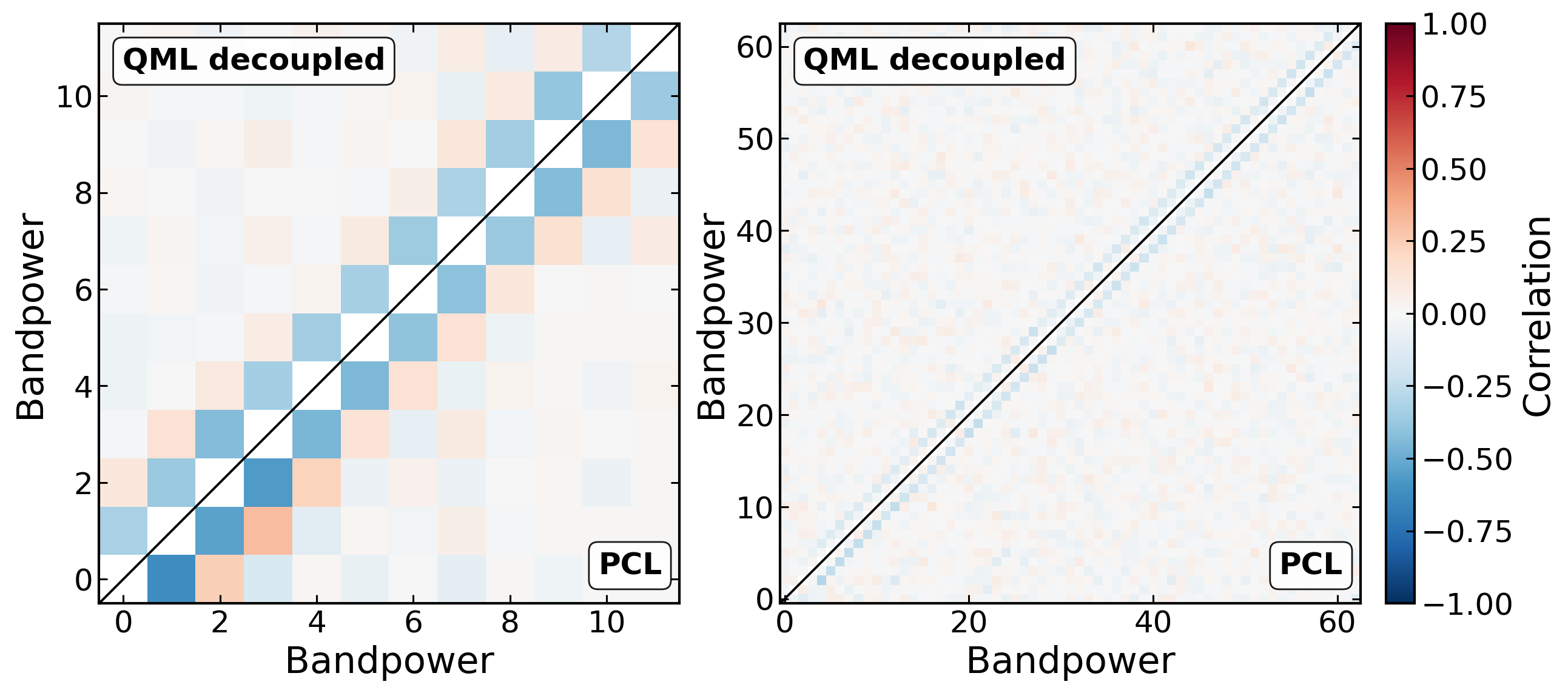}
    \caption{Empirical bandpower correlation matrices for the same configuration as Fig.~\ref{fig:qml_vs_pcl_variance}. Each panel uses a split-triangle layout: \cf's decoupled QML correlation in the upper triangle, {\tt NaMaster} decoupled pseudo-$C_\ell$ in the lower triangle, for $\fsky \approx 0.10$ (left) and $\fsky \approx 0.60$ (right). The QML correlation is consistent with diagonal at both sky fractions; the pseudo-$C_\ell$ correlation retains the off-diagonal structure inherited from mask-induced mode coupling.}
    \label{fig:qml_vs_pcl_corr}
\end{figure*}

For upcoming experiments, at LiteBIRD-relevant $\fsky$ the $\sim\!50\%$ variance inflation at the reionisation bump and recombination $B$-mode peak is not recoverable by extra integration time, while at the higher ground-based $\fsky$ of the Simons Observatory the few-to-twenty-percent gain still tightens constraints directly \citep{LiteBIRD2023, SimonsObservatory2019}. Figure~\ref{fig:qml_vs_pcl_corr} shows the same picture in the empirical bandpower correlation matrices: the QML estimate (upper triangle) is consistent with diagonal, while the pseudo-$C_\ell$ counterpart (lower triangle) retains the off-diagonal structure inherited from the mode-coupling matrix, more pronounced at $\fsky \approx 0.10$. The QML decorrelated normalisation $\hat C_\ell = \Fmat^{-1/2}\,q_\ell$ (Sect.~\ref{sec:normalisation}) is independently consistent with the identity on the same simulations (diagonal mean $0.97/0.99$ at $\fsky \approx 0.10/0.60$, off-diagonal RMS $\approx 0.03$ in both).

\section{Performance}
\label{sec:performance}

This section reports empirical wall-time behaviour of the public \cf pipeline, complementing the algorithmic derivations of Sect.~\ref{sec:formalism} and the architectural choices of Sect.~\ref{sec:architecture}. The optimisations responsible for the scaling reported here are documented in Appendix~\ref{app:performance}; we focus in this section on the measurements themselves and on what they imply for the practical operating regime of the framework.

All measurements were collected on the CINECA Galileo100 cluster (each node equipped with $2\times$ Intel Xeon Platinum 8260 ``Cascade Lake'' CPUs, 24 cores each at 2.4\,GHz, for a total of 48 cores per node and 384\,GB RAM, scipy-openblas linear algebra) from a single git revision. The benchmark scripts that produce these measurements are distributed with the codebase under \texttt{src/cosmoforge.qube/benchmarks/}. The wall-time numbers reported here cover the operating regime a user is most likely to encounter; per-stage memory characterisation is reported separately in Appendix~\ref{app:perf_memory_appendix}.

\subsection{Numba JIT acceleration}
\label{sec:perf_numba}

The two performance-critical Python kernels in \cc are the spin-0 Legendre evaluation \texttt{legendre\_plm} and the pixel-space signal matrix builder \texttt{compute\_signal\_matrix}. Both are JIT-compiled through \texttt{Numba} \citep{numba}; the first invocation pays the compilation cost, and subsequent calls run in optimised native code. Table~\ref{tab:numba_jit} reports the warmup and steady-state timings measured on a single Galileo100 node.

\begin{table}
\caption{Numba JIT warmup vs.\ steady-state cost for the two performance-critical kernels, measured on a single Galileo100 node. \texttt{legendre\_plm} is timed for a full HEALPix sphere at $\nside = 16$, $\lmax = 32$; \texttt{compute\_signal\_matrix} builds a $3200 \times 3200$ pixel-space signal covariance with $\lmaxsig = 64$.}
\label{tab:numba_jit}
\vskip -3mm
\nointerlineskip
\setbox\tablebox=\vbox{
\halign{
\hbox to 1.9in{#\leaderfil}\tabskip 1em&
\hfil#\hfil\tabskip 1.5em&
\hfil#\hfil\tabskip 0pt\cr
\noalign{\doubleline}
\omit\hfil Kernel\hfil& Warmup& Steady-state\cr
\noalign{\vskip 3pt\hrule\vskip 3pt}
\texttt{legendre\_plm}&             $0.27$\,s& $14.1$\,ms\cr
\texttt{compute\_signal\_matrix}&   $2.92$\,s& $1.23$\,s\cr
\noalign{\vskip 3pt\hrule\vskip 3pt}}}
\endPlancktable
\end{table}

The compilation overhead is paid exactly once per process: by the time a Fisher matrix or QML estimate has been produced, the kernels have been called many thousands of times and the warmup contribution is negligible compared with the linear-algebra cost. The functions are therefore safe to use inside hot loops without manual caching. The warmup-to-steady-state ratio differs sharply between the two kernels, $\sim\!19\times$ for \texttt{legendre\_plm} and $\sim\!2.4\times$ for \texttt{compute\_signal\_matrix}: the Legendre kernel is dominated by tight scalar arithmetic in which the JIT-compiled inner loop benefits maximally from native code, whereas the signal-matrix builder spends most of its time in BLAS-bound dense operations whose pre- and post-JIT cost is comparable.

\subsection{Scaling with $\nside$ and $\lmax$}
\label{sec:perf_scaling}

The most demanding regime in routine use is the small-sky-fraction limit. At $\fsky \lesssim 1\%$, even high-resolution maps have a modest active pixel count, and the pixel basis is dramatically smaller than its harmonic counterpart: at $\nside = 256$, $\lmax = 4 \nside = 1024$, the harmonic basis carries $\nmodes \sim 10^5$ modes whereas the pixel basis at $\fsky = 0.01$ carries fewer than $1.6 \times 10^4$ active pixels, an order-of-magnitude saving in the dimension that drives the cubic linear-algebra cost. Table~\ref{tab:scaling} reports total wall-time for an \texttt{auto} pipeline run (Fisher matrix plus a ten-simulation QML loop) across $\nside \in \{16, 32, 64, 128, 256\}$ in two field configurations and at two sky fractions ($\fsky = 0.01$, $\fsky = 0.1$). The selector consistently picks the pixel-direct path in this regime; near the basis crossover, where the unmodelled prefactors of the $\nmodes^3$ harmonic-SMW and $(\nbins+1)\,\npix^3$ pixel-direct cost models can be of comparable size, users with prior knowledge of the regime can override the default by passing \texttt{method="harmonic"} or \texttt{method="pixel"} (the latter with \texttt{use\_direct=True} for the no-projector pipeline characterised here).

\begin{table}
\caption{Total wall-time (Fisher + ten-simulation QML) at $\fsky = 0.01$ and $\fsky = 0.1$, with the \texttt{auto} basis selector active (it picks pixel-direct in every cell). The largest entry corresponds to a $15\,624 \times 15\,624$ dense joint $Q,U$ covariance; $\npix$ aggregates the two Stokes components for $QU$ rows.}
\label{tab:scaling}
\vskip -3mm
\nointerlineskip
\setbox\tablebox=\vbox{
\newdimen\digitwidth \setbox0=\hbox{\rm 0} \digitwidth=\wd0
\catcode`*=\active \def*{\kern\digitwidth}
\halign{
\hfil#\hfil\tabskip 1.5em&
\hfil#\tabskip 1.5em&
\hfil#\tabskip 1.5em&
\hfil#\tabskip 1.5em&
\hfil#\tabskip 0pt\cr
\noalign{\doubleline}
\omit\hfil Field\hfil& \omit\hfil $\nside$\hfil& \omit\hfil $\npix$\hfil& \omit\hfil Total (s)\hfil& \omit\hfil QML/sim (s)\hfil\cr
\noalign{\vskip 3pt\hrule\vskip 2pt}
\noalign{\vskip 1pt}\multispan5\hfil\textit{$\fsky = 0.01$}\hfil\cr\noalign{\vskip 1pt}
$T$  & *16 & ***24 & ***4.15 & *0.014\cr
$T$  & *32 & **112 & ***1.00 & *0.082\cr
$T$  & *64 & **480 & ***4.61 & *0.339\cr
$T$  & 128 & *1984 & **24.05 & *1.670\cr
$T$  & 256 & *7812 & *315.26 & 18.04*\cr
$QU$ & *16 & ***48 & ***3.02 & *0.008\cr
$QU$ & *32 & **224 & ***1.76 & *0.150\cr
$QU$ & *64 & **960 & ***8.69 & *0.512\cr
$QU$ & 128 & *3968 & **93.32 & *7.087\cr
$QU$ & 256 & 15624 & 1334.19 & 38.09*\cr
\noalign{\vskip 2pt\hrule\vskip 2pt}
\noalign{\vskip 1pt}\multispan5\hfil\textit{$\fsky = 0.1$}\hfil\cr\noalign{\vskip 1pt}
$T$  & *16 & **312 & ***4.49 & *0.030\cr
$T$  & *32 & *1200 & ***1.64 & *0.108\cr
$T$  & *64 & *4900 & **22.46 & *0.993\cr
$QU$ & *16 & **624 & ***3.90 & *0.045\cr
$QU$ & *32 & *2400 & ***8.97 & *0.260\cr
$QU$ & *64 & *9800 & *159.57 & *4.165\cr
\noalign{\vskip 3pt\hrule\vskip 3pt}}}
\endPlancktable
\end{table}

The empirical scaling with $\nside$ is consistent with the cubic dependence on $\npix$ predicted by the dense linear-algebra cost model. From $\nside = 64 \to 128 \to 256$, the active pixel count grows by approximately $4\times$ per step (the precise factor depends on the discretised mask boundary), so a strictly cubic cost would scale as $64\times$ per step. The measured factors are $8.7\,\mathrm{s} \to 93\,\mathrm{s} \to 1334\,\mathrm{s}$ in $QU$ at $\fsky = 0.01$, i.e.\ $11\times$ and $14\times$ between successive nside doublings, both well within the cubic-cost envelope after accounting for the actual mask area and for the constant-factor overheads (covariance I/O, SMW kernel assembly) that contribute a non-negligible offset at small nside. The bottom-right corner of the $\fsky=0.01$ block demonstrates that the framework reaches $\nside = 256$ with a general (non-diagonal) noise covariance within a single-node, single-allocation budget: a full Fisher matrix and ten QML simulations of a $15\,624 \times 15\,624$ joint $Q,U$ covariance complete in approximately $22$ minutes on a $48$-core node. We do not claim a record here --- ECLIPSE \citep{BilbaoAhedo2021} reports analyses at $\nside = 64$, $\lmax = 192$ on $144$ cores using its diagonal-noise specialisation, and a direct head-to-head comparison would have to be performed at matching configurations and assumptions --- but the configuration is well above the historical operating regime of the Planck low-$\ell$ reference at the same level of generality on the noise model. At $\fsky = 0.1$ the qualitative scaling is the same, with wall-times larger in proportion to the active pixel count.

This table covers the operating regime relevant to upcoming low-$\ell$ CMB analyses: $\fsky \approx 0.01$ for the deepest small-patch experiments and $\fsky \approx 0.1$ for the joint-mask configurations more typical of full-sky satellite or ground-based-survey settings. The framework reaches $\nside = 256$ in joint $Q,U$ at $\fsky = 0.01$ within a single-node, single-allocation budget. In this regime the pixel basis is decisively cheaper than the harmonic alternative, and the \texttt{auto} selector picks it in every cell without user intervention.

\subsection{MPI strong scaling}
\label{sec:perf_mpi}

\cf parallelises three independent stages over MPI ranks (Appendix~\ref{app:implementation}): Fisher trace contractions, simulation-by-simulation QML evaluation in \texttt{Spectra}, and parameter-grid sweeps in \pics. We characterise the first two on the same fixed problem ($QU$, $\nside=32$, $\lmax=64$, $10^4$ simulations) at fixed total core budget, sweeping rank count $N \in \{1,2,4,8\}$ with the per-rank thread budget adjusted as $N_\mathrm{OMP} = 48/N$. Table~\ref{tab:mpi_scaling} reports the wall-time per stage.

\begin{table}
\caption{Hybrid MPI$\,\times\,$OpenMP scaling at fixed $N \times N_\mathrm{OMP} = 48$ on a single Galileo100 node. Configuration: $QU$, $\nside=32$, $\lmax=64$, harmonic basis, $10^4$ Gaussian simulations. ``Fisher'' is the full pipeline cost (setup + traces); ``Spectra'' is the per-batch wall-time for the full simulation set.}
\label{tab:mpi_scaling}
\vskip -3mm
\nointerlineskip
\setbox\tablebox=\vbox{
\newdimen\digitwidth \setbox0=\hbox{\rm 0} \digitwidth=\wd0
\catcode`*=\active \def*{\kern\digitwidth}
\halign{
\hfil#\tabskip 1.5em&
\hfil#\tabskip 1.5em&
\hfil#\tabskip 1.5em&
\hfil#\tabskip 1.5em&
\hfil#\tabskip 0pt\cr
\noalign{\doubleline}
\omit\hfil $N$\hfil& \omit\hfil $N_\mathrm{OMP}$\hfil& \omit\hfil Fisher (s)\hfil& \omit\hfil Spectra (s)\hfil& \omit\hfil QML/sim (ms)\hfil\cr
\noalign{\vskip 3pt\hrule\vskip 3pt}
1 & 48 & $197.9$ & $*9.16$ & $0.92$\cr
2 & 24 & $230.7$ & $13.35$ & $1.34$\cr
4 & 12 & $268.5$ & $21.05$ & $2.11$\cr
8 & *6 & $375.2$ & $38.91$ & $3.89$\cr
\noalign{\vskip 3pt\hrule\vskip 3pt}}}
\endPlancktable
\end{table}

Both stages \emph{anti-scale} with $N$ at fixed total core budget: the BLAS-dominated rank-0 setup work (V build, effective-noise inversion, SMW kernel assembly) is single-process and benefits directly from the per-rank thread budget, so subdividing $48$ cores into $48/N$-thread slices linearly stretches it; \texttt{Spectra} follows the same pattern through its per-simulation BLAS calls. A reference point at $N\!=\!1$, $N_\mathrm{OMP}\!=\!1$ returns Fisher $1156$\,s and Spectra $152$\,s --- $\sim\!6\times$ slower than the $N\!=\!1$, $N_\mathrm{OMP}\!=\!48$ row --- confirming the workload is BLAS-thread-bound. Cross-rank consistency on the recovered Fisher matrix and QML estimates is at machine precision ($\sim\!10^{-16}$) across all configurations. The MPI distribution does deliver near-linear speedups where the parallel work is not bottlenecked by single-process setup --- \pics's parameter-grid scan and cross-spectrum loops over independent map pairs are both distributed round-robin across ranks with no shared single-process stage (the parameter-grid scan of Sect.~\ref{sec:picslike_inference} is run in this configuration). An alternative MPI partitioning that distributes the SMW kernel assembly and the per-bin trace stage across ranks is a natural target for follow-up work.

Deploy with one MPI rank per node and OpenMP threads matched to the available cores; the default thread placement is reliable (explicit \texttt{OMP\_PROC\_BIND=spread} degrades single-process BLAS in our tests). Per-node memory sizing follows the closed-form persistent-state model and calibrated peak-RSS sweep of Appendix~\ref{app:perf_memory_appendix}, within a single-node budget for the configurations characterised here.


\section{Conclusions and outlook}
\label{sec:conclusions}

We have presented \cf, a Python framework that integrates QML power spectrum estimation and pixel-space Gaussian likelihood evaluation for spin-0 and spin-2 fields with general (non-diagonal) noise covariances, exposing a harmonic-space SMW path and a direct pixel-space path through a single abstract computation-basis interface (Sect.~\ref{sec:harmonic_basis_code}). The implementation is validated end-to-end against the Planck low-$\ell$ Fortran reference of \citet{Pagano2020} and the \texttt{simall} likelihood of \citet{PlanckV2020} and exercised on $10^{4}$ Monte Carlo polarisation realisations across five sky fractions (Sect.~\ref{sec:fortran_validation}). Within the QML code landscape (Appendix~\ref{app:comparison}), \cf is the only public framework that simultaneously exposes three normalisation modes, both computation bases through a single abstract interface, an integrated pixel-space likelihood, and a per-component multipole floor, all in Python.

Several extensions of the released framework are at various stages of development and will be presented in dedicated follow-up papers. Iterative QML \citep{BilbaoAhedo2021} can be driven on top of \cf by re-entering the pipeline with the previous-iteration estimates in place of the fiducial, leveraging the modular separation between fiducial ingestion and basis assembly; an end-to-end pixel-based-inference characterisation, including application to real CMB data and integration with cosmological samplers, is the subject of a forthcoming companion paper based on the demonstration in Sect.~\ref{sec:picslike_inference}; and GPU acceleration of the pixel-space hot loops via \texttt{CuPy} or \texttt{JAX} is localised to the basis kernels by the abstract interface and is the obvious next step where the dense linear algebra dominates. The framework targets the regime of upcoming low-$\ell$ CMB analyses --- LiteBIRD \citep{LiteBIRD2023} and the Simons Observatory \citep{SimonsObservatory2019} --- where correlated noise, partial sky, and a primordial $B$-mode target make optimal estimators essential, and extends to non-CMB spherical observables (galaxy clustering, 21\,cm intensity mapping, stochastic GWB searches) and their cross-correlations through the multi-field design.

\cf replaces the historical proliferation of single-purpose QML codes with a single, modular, validated, and openly-developed framework, lowering the barrier to entry for optimal large-scale analyses and providing a reliable foundation on which future algorithmic and scientific work can build.

\cf is publicly available at \url{https://github.com/ggalloni/CosmoForge}; the three subpackages \cc, \qube, \pics\ are pip-installable from PyPI as \texttt{cosmocore}, \texttt{qube-qml}, \texttt{picslike}, and the umbrella \texttt{cosmoforge} installs all three.

\begin{acknowledgements}
The authors thank L.P.L.\ Colombo for valuable comments and discussions. Some of the results in this paper have been derived using the following packages: {\tt CAMB} \citep{Lewis2000, Howlett2012}, {\tt healpy} \citep{Gorski2005, Zonca2019}, {\tt Matplotlib} \citep{Hunter2007}, {\tt SciPy} \citep{Virtanen2020}, {\tt NumPy} \citep{Harris2020}, {\tt Numba} \citep{numba}, and {\tt mpi4py} \citep{mpi4py}. We acknowledge the financial support from the INFN InDark initiative and from the COSMOS network through the ASI (Italian Space Agency) Grants 2016-24-H.0 and 2016-24-H.1-2018. This work has also received funding by the European Union's Horizon 2020 research and innovation program under grant agreement no.\ 101007633 CMB-Inflate. G.G.\ acknowledges support by the MUR PRIN2022 Project ``BROWSEPOL: Beyond standaRd mOdel With coSmic microwavE background POLarization''-2022EJNZ53 financed by the European Union -- Next Generation EU. We acknowledge CINECA for the availability of high performance computing resources and support through the CINECA-INFN agreement. During the preparation of this work the authors used Anthropic's Claude (via the Claude Code agentic coding interface) to assist with code implementation and manuscript drafting. All AI-generated content and code was reviewed, edited, and validated by the authors, who take full responsibility for the contents of this publication.
\end{acknowledgements}


\bibliographystyle{bibtex/aa}
\bibliography{references}

@ARTICLE{Tegmark1997,
    author  = {Tegmark, M.},
    title   = {How to measure {CMB} power spectra without losing information},
    journal = {Phys.\ Rev.\ D},
    volume  = {55},
    pages   = {5895},
    year    = {1997}
}

@ARTICLE{TegmarkOliveiraCosta2001,
    author  = {Tegmark, M. and de Oliveira-Costa, A.},
    title   = {How to measure {CMB} polarization power spectra without losing information},
    journal = {Phys.\ Rev.\ D},
    volume  = {64},
    pages   = {063001},
    year    = {2001}
}

@INPROCEEDINGS{TegmarkHamilton1997,
    author        = {Tegmark, M. and Hamilton, A. J. S.},
    title         = {Uncorrelated measurements of the {CMB} power spectrum},
    booktitle     = {Relativistic Astrophysics: 18th Texas Symposium},
    year          = {1997},
    eprint        = {astro-ph/9702019},
    archivePrefix = {arXiv}
}

@ARTICLE{BondJaffeKnox1998,
    author  = {Bond, J. R. and Jaffe, A. H. and Knox, L.},
    title   = {Estimating the power spectrum of the cosmic microwave background},
    journal = {Phys.\ Rev.\ D},
    volume  = {57},
    pages   = {2117},
    year    = {1998}
}

@ARTICLE{Kamionkowski1997,
    author  = {Kamionkowski, M. and Kosowsky, A. and Stebbins, A.},
    title   = {Statistics of cosmic microwave background polarization},
    journal = {Phys.\ Rev.\ D},
    volume  = {55},
    pages   = {7368},
    year    = {1997}
}

@ARTICLE{ZaldarriagaSeljak1997,
    author  = {Zaldarriaga, M. and Seljak, U.},
    title   = {All-sky analysis of polarization in the microwave background},
    journal = {Phys.\ Rev.\ D},
    volume  = {55},
    pages   = {1830},
    year    = {1997}
}

@ARTICLE{Goldberg1967,
    author  = {Goldberg, J. N. and Macfarlane, A. J. and Newman, E. T. and Rohrlich, F. and Sudarshan, E. C. G.},
    title   = {Spin-$s$ spherical harmonics and {$\eth$}},
    journal = {J.\ Math.\ Phys.},
    volume  = {8},
    pages   = {2155},
    year    = {1967}
}

@TECHREPORT{Woodbury1950,
    author      = {Woodbury, M. A.},
    title       = {Inverting modified matrices},
    institution = {Memorandum Report 42, Statistical Research Group, Princeton University},
    year        = {1950}
}

@BOOK{GolubVanLoan2013,
    author    = {Golub, G. H. and Van Loan, C. F.},
    title     = {Matrix Computations},
    edition   = {4},
    publisher = {Johns Hopkins University Press},
    year      = {2013}
}

@ARTICLE{BilbaoAhedo2021,
    author        = {Bilbao-Ahedo, J. D. and Barreiro, R. B. and Vielva, P. and Mart{\'\i}nez-Gonz{\'a}lez, E. and Herranz, D.},
    title         = {{ECLIPSE}: a fast Quadratic Maximum Likelihood estimator for {CMB} intensity and polarization power spectra},
    journal       = {JCAP},
    volume        = {2021},
    number        = {07},
    pages         = {034},
    year          = {2021},
    eprint        = {2104.08528},
    archivePrefix = {arXiv},
    doi           = {10.1088/1475-7516/2021/07/034}
}

@ARTICLE{Vanneste2018,
    author        = {Vanneste, S. and Henrot-Versill{\'e}, S. and Louis, T. and Tristram, M.},
    title         = {Quadratic estimator for {CMB} cross-correlation},
    journal       = {Phys.\ Rev.\ D},
    volume        = {98},
    pages         = {103526},
    year          = {2018},
    eprint        = {1807.02484},
    archivePrefix = {arXiv}
}

@ARTICLE{Gruppuso2009,
    author  = {Gruppuso, A. and de Rosa, A. and Cabella, P. and Paci, F. and Finelli, F. and Natoli, P. and de Gasperis, G. and Mandolesi, N.},
    title   = {New estimates of the {CMB} angular power spectra from the {WMAP} 5-year low-resolution data},
    journal = {MNRAS},
    volume  = {400},
    pages   = {463},
    year    = {2009},
    eprint  = {0904.0789},
    archivePrefix = {arXiv},
    doi     = {10.1111/j.1365-2966.2009.15469.x}
}

@ARTICLE{Gjerlow2015,
    author        = {Gjerl{\o}w, E. and Colombo, L. P. L. and Eriksen, H. K. and G{\'o}rski, K. M. and Gruppuso, A. and Jewell, J. B. and Plaszczynski, S. and Wehus, I. K.},
    title         = {Optimized large-scale {CMB} likelihood and quadratic maximum likelihood power spectrum estimation},
    journal       = {ApJS},
    volume        = {221},
    pages         = {5},
    year          = {2015},
    eprint        = {1506.04273},
    archivePrefix = {arXiv},
    doi           = {10.1088/0067-0049/221/1/5}
}

@ARTICLE{Kvasiuk2025,
    author        = {Kvasiuk, Y. and Lai, A. and M{\"u}nchmeyer, M. and Smith, K. M.},
    title         = {{QML-FAST}: a fast code for low-{$\ell$} tomographic maximum likelihood power spectrum estimation},
    journal       = {arXiv e-prints},
    year          = {2025},
    eprint        = {2510.05215},
    archivePrefix = {arXiv}
}

@ARTICLE{Hivon2002,
    author  = {Hivon, E. and G{\'o}rski, K. M. and Netterfield, C. B. and Crill, B. P. and Prunet, S. and Hansen, F.},
    title   = {{MASTER} of the cosmic microwave background anisotropy power spectrum: a fast method for statistical analysis of large and complex cosmic microwave background data sets},
    journal = {ApJ},
    volume  = {567},
    pages   = {2},
    year    = {2002}
}

@ARTICLE{Alonso2019,
    author  = {Alonso, D. and Sanchez, J. and Slosar, A. and {LSST Dark Energy Science Collaboration}},
    title   = {A unified pseudo-{$C_\ell$} framework},
    journal = {MNRAS},
    volume  = {484},
    pages   = {4127},
    year    = {2019},
    eprint  = {1809.09603},
    archivePrefix = {arXiv},
    doi     = {10.1093/mnras/stz093}
}

@ARTICLE{Gorski2005,
    author  = {G{\'o}rski, K. M. and Hivon, E. and Banday, A. J. and Wandelt, B. D. and Hansen, F. K. and Reinecke, M. and Bartelmann, M.},
    title   = {{HEALPix}: a framework for high-resolution discretization and fast analysis of data distributed on the sphere},
    journal = {ApJ},
    volume  = {622},
    pages   = {759},
    year    = {2005}
}

@ARTICLE{Lewis2000,
    author  = {Lewis, A. and Challinor, A. and Lasenby, A.},
    title   = {Efficient computation of {CMB} anisotropies in closed {FRW} models},
    journal = {ApJ},
    volume  = {538},
    pages   = {473},
    year    = {2000},
    eprint  = {astro-ph/9911177},
    archivePrefix = {arXiv}
}

@ARTICLE{Howlett2012,
    author  = {Howlett, C. and Lewis, A. and Hall, A. and Challinor, A.},
    title   = {{CMB} power spectrum parameter degeneracies in the era of precision cosmology},
    journal = {JCAP},
    volume  = {2012},
    number  = {04},
    pages   = {027},
    year    = {2012},
    eprint  = {1201.3654},
    archivePrefix = {arXiv}
}

@ARTICLE{Zonca2019,
    author  = {Zonca, A. and Singer, L. and Lenz, D. and Reinecke, M. and Rosset, C. and Hivon, E. and Gorski, K.},
    title   = {{healpy}: equal area pixelization and spherical harmonics transforms for data on the sphere in {Python}},
    journal = {J.\ Open Source Softw.},
    volume  = {4},
    pages   = {1298},
    year    = {2019}
}

@ARTICLE{Hunter2007,
    author  = {Hunter, J. D.},
    title   = {{Matplotlib}: a {2D} graphics environment},
    journal = {Comput.\ Sci.\ Eng.},
    volume  = {9},
    pages   = {90},
    year    = {2007}
}

@ARTICLE{Virtanen2020,
    author  = {Virtanen, P. and Gommers, R. and Oliphant, T. E. and others},
    title   = {{SciPy} 1.0: fundamental algorithms for scientific computing in {Python}},
    journal = {Nat.\ Methods},
    volume  = {17},
    pages   = {261},
    year    = {2020}
}

@ARTICLE{Harris2020,
    author  = {Harris, C. R. and Millman, K. J. and van der Walt, S. J. and others},
    title   = {Array programming with {NumPy}},
    journal = {Nature},
    volume  = {585},
    pages   = {357},
    year    = {2020}
}

@ARTICLE{mpi4py,
    author  = {Dalcin, L. and Fang, Y.-L. L.},
    title   = {{mpi4py}: Status Update After 12 Years of Development},
    journal = {Comput. Sci. Eng.},
    volume  = {23},
    number  = {4},
    pages   = {47},
    year    = {2021},
    doi     = {10.1109/MCSE.2021.3083216}
}

@INPROCEEDINGS{numba,
    author    = {Lam, Siu Kwan and Pitrou, Antoine and Seibert, Stanley},
    title     = {{Numba}: a {LLVM}-based {Python} {JIT} compiler},
    booktitle = {Proceedings of the Second Workshop on the {LLVM} Compiler Infrastructure in {HPC}},
    series    = {LLVM '15},
    year      = {2015},
    publisher = {ACM}
}

@ARTICLE{PlanckXI2016,
    author        = {{Planck Collaboration}},
    title         = {{Planck} 2015 results.\ {XI}.\ {CMB} power spectra, likelihoods, and robustness of parameters},
    journal       = {A\&A},
    volume        = {594},
    pages         = {A11},
    year          = {2016},
    eprint        = {1507.02704},
    archivePrefix = {arXiv}
}

@ARTICLE{Benabed2009,
    author  = {Benabed, K. and Cardoso, J.-F. and Prunet, S. and Hivon, E.},
    title   = {{TEASING} parity-violating {B}-modes},
    journal = {A\&A},
    volume  = {500},
    pages   = {969},
    year    = {2009}
}

@ARTICLE{Aghanim2020likelihood,
    author        = {{Planck Collaboration} and Aghanim, N. and others},
    title         = {{Planck 2018 results. V. CMB power spectra and likelihoods}},
    journal       = {A\&A},
    volume        = {641},
    pages         = {A5},
    year          = {2020},
    eprint        = {1907.12875},
    archivePrefix = {arXiv}
}

@ARTICLE{Akrami2020npipe,
    author        = {{Planck Collaboration} and Akrami, Y. and others},
    title         = {{Planck intermediate results. LVII. Joint Planck LFI and HFI data processing}},
    journal       = {A\&A},
    volume        = {643},
    pages         = {A42},
    year          = {2020},
    eprint        = {2007.04997},
    archivePrefix = {arXiv}
}

@ARTICLE{PlanckXLVI2016,
    author        = {{Planck Collaboration}},
    title         = {{Planck} intermediate results.\ {XLVI}.\ Reduction of large-scale systematic effects in {HFI} polarization maps and estimation of the reionization optical depth},
    journal       = {A\&A},
    volume        = {596},
    pages         = {A107},
    year          = {2016},
    eprint        = {1605.02985},
    archivePrefix = {arXiv}
}

@ARTICLE{PlanckV2020,
    author  = {{Planck Collaboration}},
    title   = {{Planck} 2018 results.\ {V}.\ {CMB} power spectra and likelihoods},
    journal = {A\&A},
    volume  = {641},
    pages   = {A5},
    year    = {2020}
}

@ARTICLE{Pagano2020,
    author        = {Pagano, L. and Delouis, J.-M. and Mottet, S. and Puget, J.-L. and Vibert, L.},
    title         = {Reionization optical depth determination from {Planck} {HFI} data with ten percent accuracy},
    journal       = {A\&A},
    volume        = {635},
    pages         = {A99},
    year          = {2020},
    eprint        = {1908.09856},
    archivePrefix = {arXiv}
}

@ARTICLE{BICEPKeck2021,
    author  = {{BICEP/Keck Collaboration}},
    title   = {Improved constraints on primordial gravitational waves using {Planck}, {WMAP}, and {BICEP/Keck} observations through the 2018 observing season},
    journal = {Phys.\ Rev.\ Lett.},
    volume  = {127},
    pages   = {151301},
    year    = {2021}
}

@ARTICLE{Tristram2022,
    author  = {Tristram, M. and others},
    title   = {Improved limits on the tensor-to-scalar ratio using {BICEP} and {Planck} data},
    journal = {Phys.\ Rev.\ D},
    volume  = {105},
    pages   = {083524},
    year    = {2022}
}

@ARTICLE{LiteBIRD2023,
    author  = {{LiteBIRD Collaboration}},
    title   = {Probing cosmic inflation with the {LiteBIRD} cosmic microwave background polarization survey},
    journal = {PTEP},
    volume  = {2023},
    pages   = {042F01},
    year    = {2023}
}

@ARTICLE{SimonsObservatory2019,
    author  = {Ade, P. and others},
    title   = {The {Simons Observatory}: science goals and forecasts},
    journal = {JCAP},
    volume  = {02},
    pages   = {056},
    year    = {2019}
}

@ARTICLE{Galloni2025,
    author       = {Galloni, G. and Campeti, P. and Pagano, L. and Gerbino, M. and Lattanzi, M. and Natoli, P.},
    title        = {Accurate and efficient likelihood modeling for large-scale {CMB} data},
    journal      = {JCAP},
    volume       = {12},
    pages        = {052},
    year         = {2025},
    eprint       = {2505.24829},
    archivePrefix = {arXiv}
}

@ARTICLE{Chon2004,
    author  = {Chon, G. and Challinor, A. and Prunet, S. and Hivon, E. and Szapudi, I.},
    title   = {Fast estimation of polarization power spectra using correlation functions},
    journal = {MNRAS},
    volume  = {350},
    pages   = {914},
    year    = {2004},
    doi     = {10.1111/j.1365-2966.2004.07737.x}
}

@ARTICLE{Tristram2005,
    author  = {Tristram, M. and Mac{\'\i}as-P{\'e}rez, J. F. and Renault, C. and Santos, D.},
    title   = {{XSPECT}, estimation of the angular power spectrum by computing cross-power spectra with analytical error bars},
    journal = {MNRAS},
    volume  = {358},
    pages   = {833},
    year    = {2005},
    doi     = {10.1111/j.1365-2966.2005.08760.x}
}

@ARTICLE{Grain2009,
    author  = {Grain, J. and Tristram, M. and Stompor, R.},
    title   = {Polarized {CMB} power spectrum estimation using the pure pseudo-cross-spectrum approach},
    journal = {PhRvD},
    volume  = {79},
    pages   = {123515},
    year    = {2009},
    doi     = {10.1103/PhysRevD.79.123515},
    eprint  = {0903.2350},
    archivePrefix = {arXiv}
}

@ARTICLE{Page2007,
    author  = {Page, L. and Hinshaw, G. and Komatsu, E. and Nolta, M. R. and Spergel, D. N. and Bennett, C. L. and Barnes, C. and Bean, R. and Dor\'e, O. and Dunkley, J. and Halpern, M. and Hill, R. S. and Jarosik, N. and Kogut, A. and Limon, M. and Meyer, S. S. and Odegard, N. and Peiris, H. V. and Tucker, G. S. and Verde, L. and Weiland, J. L. and Wollack, E. and Wright, E. L.},
    title   = {Three-Year {Wilkinson Microwave Anisotropy Probe (WMAP)} Observations: Polarization Analysis},
    journal = {ApJS},
    volume  = {170},
    pages   = {335},
    year    = {2007},
    doi     = {10.1086/513699},
    eprint  = {astro-ph/0603450},
    archivePrefix = {arXiv}
}

@ARTICLE{Jensen1906,
    author  = {Jensen, J. L. W. V.},
    title   = {{Sur les fonctions convexes et les in\'egalit\'es entre les valeurs moyennes}},
    journal = {Acta Mathematica},
    volume  = {30},
    pages   = {175},
    year    = {1906},
    doi     = {10.1007/BF02418571}
}

\begin{appendix}
\nolinenumbers  


\section{Spin-weighted spherical harmonics and rotation angles}
\label{app:spin_harmonics}

\subsection{Spin-2 extension}
\label{sec:spin2}

The formalism presented so far applies to spin-0 (scalar) fields such as CMB temperature or galaxy density. Spin-2 (tensor) fields --- most notably CMB polarization, described by the Stokes parameters $Q$ and $U$, and weak-lensing galaxy shear $(\gamma_1, \gamma_2)$ --- require a generalisation to spin-weighted spherical harmonics. In this section we describe how each ingredient of the QML estimator is modified for spin-2 fields and for cross-correlations between spin-0 and spin-2 fields.

\subsubsection{Spin-weighted harmonics and the V operator}
\label{sec:spin2_V}

A spin-2 field on the sphere is decomposed into $E$-mode (parity-even) and $B$-mode (parity-odd) components using the spin-weighted spherical harmonics ${}_{\pm 2}Y_{\ell m}$ \citep{Kamionkowski1997, ZaldarriagaSeljak1997}. In pixel space, a spin-2 field contributes $2\npix^\mathrm{phys}$ data points (the $Q$ and $U$ values at each unmasked pixel), while in harmonic space it contributes $2 \nmodes^\mathrm{base}$ modes ($E$ and $B$ modes for each $(\ell, m)$).

The harmonic operator $\Vmat$ for a spin-2 field has the block structure
\begin{equation}
    \Vmat_\mathrm{spin\text{-}2} =
    \begin{pmatrix}
        \Vmat^{EQ} & \Vmat^{EU} \\
        \Vmat^{BQ} & \Vmat^{BU}
    \end{pmatrix},
    \label{eq:V_spin2}
\end{equation}
where rows correspond to $[E\text{ modes}\,|\,B\text{ modes}]$ and columns to $[Q\text{ pixels}\,|\,U\text{ pixels}]$. The sub-blocks are constructed from Wigner $d$-matrix elements $d^\ell_{m,\pm 2}(\theta)$, evaluated via a numerically stable three-term recurrence in $\ell$ derived from the Jacobi polynomial recurrence. Rotation angles between pixel coordinate frames, computed from the full spherical geometry \citep[not a flat-sky approximation; see][]{Kamionkowski1997}, are used to properly transform between the local $(Q,U)$ basis at each pixel and the global $(E,B)$ basis.

\subsubsection{Lambda matrix for spin-2}
\label{sec:spin2_lambda}

For a spin-0 field, $\Lmat$ is diagonal with entries $\Cl\,\Wl$. For a spin-2 field, $\Lmat$ acquires a $2 \times 2$ block structure at each $(\ell, m)$:
\begin{equation}
    \Lmat_{\ell m}^\mathrm{spin\text{-}2} =
    \begin{pmatrix}
        C_\ell^{EE}\, \Wl & C_\ell^{EB}\, \Wl \\
        C_\ell^{EB}\, \Wl & C_\ell^{BB}\, \Wl
    \end{pmatrix}.
    \label{eq:Lambda_spin2}
\end{equation}
This block-diagonal structure (diagonal in $(\ell,m)$ but $2 \times 2$ in $E/B$) is a consequence of statistical isotropy: different $(\ell,m)$ are uncorrelated, but $E$ and $B$ modes at the same $(\ell,m)$ may be correlated if $C_\ell^{EB} \neq 0$.

\subsubsection{Multi-field structure}
\label{sec:multi_field}

For a joint analysis of multiple fields --- for example, temperature ($T$, spin-0) and polarization ($Q/U$, spin-2) --- the data vector is assembled as $\dvec = (T_1, \ldots, T_{\npix^T}, Q_1, \ldots, Q_{\npix^P}, U_1, \ldots, U_{\npix^P})$, and the harmonic operator becomes block-diagonal across field components:
\begin{equation}
    \Vmat = \begin{pmatrix}
        \Vmat_T & 0 \\
        0 & \Vmat_{QU}
    \end{pmatrix}.
    \label{eq:V_multifield}
\end{equation}
The $\Lmat$ matrix acquires the full block structure coupling all spectra. For a $TEB$ analysis, each $(\ell,m)$ block is $3 \times 3$:
\begin{equation}
    \Lmat_{\ell m}^{TEB} = \Wl
    \begin{pmatrix}
        C_\ell^{TT} & C_\ell^{TE} & C_\ell^{TB} \\
        C_\ell^{TE} & C_\ell^{EE} & C_\ell^{EB} \\
        C_\ell^{TB} & C_\ell^{EB} & C_\ell^{BB}
    \end{pmatrix},
    \label{eq:Lambda_TEB}
\end{equation}
encoding all six auto- and cross-power spectra. The derivative matrices $\Emat_\ell$ and the Fisher matrix generalise accordingly, with indices running over all $(spectrum, \ell)$ pairs.

The SMW machinery (Sect.~\ref{sec:smw}) carries over without modification: $\Kmat = \Lmat^{-1} + \Vmat\Nmat^{-1}\Vmat^\top$ is now $\nmodes^\mathrm{total} \times \nmodes^\mathrm{total}$, with $\nmodes^\mathrm{total} = \nmodes^T + 2\nmodes^P$ for a $TEB$ analysis.

\paragraph{Field block-diagonal optimisation.} When no cross-spectra are estimated between certain field groups and their noise is independent, the kernel $\Kmat$ is exactly block-diagonal across those groups. \cf detects this structure automatically and inverts each block independently, reducing the cost from $\mathcal{O}((\nmodes^\mathrm{total})^3)$ to $\sum_g \mathcal{O}((\nmodes^g)^3)$. For a $TEB$ analysis without $TE$/$TB$ cross-spectra, this gives two blocks ($T$ and $EB$) and a speedup of approximately $4\times$ on the $\Kmat$ inversion.

\bigskip

The spin-2 machinery of Sect.~\ref{sec:spin2} relies on a small number of explicit conventions for the spin-weighted spherical harmonics ${}_{\pm 2} Y_{\ell m}$ and for the rotation that aligns the local Stokes frame at every pixel with the global $E$/$B$ basis. Because different communities adopt different sign conventions for these objects, and because mistakes in either of them propagate silently into the spin-0$\times$spin-2 cross-spectra, this appendix records the conventions used by \cf in a form that is sufficient for an independent implementation to reproduce our results.

\subsection{Conventions for ${}_{\pm 2}Y_{\ell m}$}
\label{app:spin_harm_conventions}

We follow the spin-raising/lowering conventions of \citet{Goldberg1967} as adapted to the CMB context by \citet{Kamionkowski1997} and \citet{ZaldarriagaSeljak1997}. The spin-weighted harmonics ${}_{s} Y_{\ell m}$ admit the explicit form
\begin{equation}
    {}_{s} Y_{\ell m}(\theta, \phi)
    = \sqrt{\frac{2\ell+1}{4\pi}}\;
      d^\ell_{m,-s}(\theta)\, e^{i m \phi},
    \label{eq:spin_y_def}
\end{equation}
where $d^\ell_{m,s}(\theta)$ is the small Wigner $d$-matrix (Sect.~\ref{app:wigner_recurrence}). The $E$ and $B$ basis functions follow as real-valued combinations of the ${}_{\pm 2} Y_{\ell m}$:
\begin{equation}
    E_{\ell m} = -\frac{1}{2}\,\bigl({}_{2} Y_{\ell m} + {}_{-2} Y_{\ell m}\bigr),
    \qquad
    B_{\ell m} = +\frac{i}{2}\,\bigl({}_{2} Y_{\ell m} - {}_{-2} Y_{\ell m}\bigr).
    \label{eq:EB_from_spin}
\end{equation}
The minus sign in the $E$ combination is the one that propagates into the spin-0$\times$spin-2 entries of $\Lmat$ (Eq.~\ref{eq:Lambda_spin2}) and into the corresponding derivative matrices. Without it, the $TE$ and $TB$ blocks of $\Vmat^\top \Lmat\, \Vmat$ acquire the wrong sign relative to the pixel-space $T$--$Q,U$ correlations, while the $TT$, $EE$, $BB$, and $EB$ blocks --- which are all even in the $\pm 2$ sum --- remain unaffected and cannot diagnose the mistake. We mention this explicitly because it is the sign convention against which our spin-0$\times$spin-2 unit tests are calibrated.

\subsection{Wigner-$d$ recurrence}
\label{app:wigner_recurrence}

The values of $d^\ell_{m,s}(\theta)$ at the pixel colatitudes are evaluated through a stable three-term recurrence in $\ell$, derived from the Jacobi polynomial three-term recurrence and implemented in \texttt{cosmocore.basics.wigner}. The recurrence takes the schematic form
\begin{equation}
    d^{\ell+1}_{m,s}(\theta) = \bigl[\alpha_\ell\,\cos\theta + \beta_\ell\bigr]\,
                               d^\ell_{m,s}(\theta)
                             + \gamma_\ell\, d^{\ell-1}_{m,s}(\theta),
    \label{eq:wigner_recurrence}
\end{equation}
with $\alpha_\ell, \beta_\ell, \gamma_\ell$ the standard Jacobi coefficients (omitted here for brevity; the explicit form is given in the source).

Two book-keeping points are worth recording. First, the recurrence is started from the boundary case $\ell_\mathrm{start} = \max(|m|, |s|)$, at which $d^{\ell_\mathrm{start}}_{m,s}$ has a closed-form expression in half-angles $\cos(\theta/2)$, $\sin(\theta/2)$. The factorials that appear in the boundary expression are evaluated through their logarithms to avoid overflow at high $\ell$. Second, the symmetry $d^\ell_{m,s} = (-1)^{m-s}\, d^\ell_{s,m} = (-1)^{m-s}\, d^\ell_{-m,-s}$ is used to remap any requested $(m, s)$ into the range $|m| \ge |s|,\; m \ge 0$ before running the recurrence; this is the regime in which Eq.~\eqref{eq:wigner_recurrence} is numerically well-behaved across the full range of $\ell$ relevant to low-$\ell$ analyses.

\subsection{Rotation angles between pixel frames}
\label{app:rotation_angles}

The Stokes parameters $Q$ and $U$ at a pixel $p$ are defined relative to the local meridian through that pixel. To express $Q_p, U_p$ as a linear combination of the global $E_{\ell m}, B_{\ell m}$ harmonic coefficients, the local frame must be rotated to coincide with the $E$/$B$ basis defined globally at the pixel position. Equivalently, the spin-2 entries of the $\Vmat$ operator (Sect.~\ref{sec:spin2_V}) carry phases $e^{\pm 2 i \alpha_p}$ that depend on the pixel direction $\hat n_p$ and on the global frame chosen for $E,B$.

For the pair-wise object $\Vmat \Vmat^\top$ that enters the projected covariance, the relevant quantity is not a single rotation angle per pixel but the pair $(\alpha_{pp'}, \alpha_{p'p})$ of angles attached to the geodesic connecting two pixels $p, p'$ to the chosen reference. \cf evaluates these angles in closed form from the unit vectors $\hat n_p, \hat n_{p'}$ as implemented in \texttt{cosmocore.basics.geometry.get\_rotation\_angle}: the cross product $\hat n_p \times \hat n_{p'}$ defines the geodesic plane, and the projection of this plane onto the local tangent plane at each pixel yields $\alpha_{pp'}, \alpha_{p'p}$ via two well-defined arc-cosines whose signs are fixed by the orientation of the geodesic relative to the reference axis. We use the full-sky spherical-trigonometric expression rather than the flat-sky approximation, so the rotation angles remain accurate at all separations on the sphere; this is required for the QML computation at low $\nside$ where pixel separations span a large fraction of $\pi$.


\section{Sherman--Morrison--Woodbury for log-determinants}
\label{app:smw_logdet}

Equation~\eqref{eq:logdet} expresses the log-determinant of the total covariance $\Cmat$ as a sum of three contributions defined on much smaller matrices than $\Cmat$ itself. The identity is a direct consequence of the matrix determinant lemma, which is less commonly stated than the Sherman--Morrison--Woodbury identity for the inverse. This appendix records the derivation, in the form in which it applies to the \cf factorisation, and the numerical considerations that follow from it.

\subsection{Matrix determinant lemma}
\label{app:mdl}

For any invertible $\Nmat \in \mathbb{R}^{\npix \times \npix}$ and rectangular $\Vmat \in \mathbb{R}^{\nmodes \times \npix}$ with $\nmodes \le \npix$, and any invertible $\Lmat \in \mathbb{R}^{\nmodes \times \nmodes}$,
\begin{equation}
    \left| \Nmat + \Vmat^\top \Lmat\, \Vmat \right|
    = |\Nmat|\, \left| \Imat_{\nmodes} + \Lmat\, \Vmat\, \Nmat^{-1}\, \Vmat^\top \right|.
    \label{eq:mdl}
\end{equation}
A short proof follows from a Schur-complement argument applied to the block matrix
\begin{equation}
    \begin{pmatrix} \Nmat & -\Vmat^\top \Lmat \\ \Vmat & \Imat_{\nmodes} \end{pmatrix}.
\end{equation}
Computing its determinant by eliminating the upper-right block via the lower-left block gives $|\Nmat|\, |\Imat_{\nmodes} + \Lmat\, \Vmat\, \Nmat^{-1}\, \Vmat^\top|$, while eliminating the lower-left block via the upper-left block gives $|\Imat_{\nmodes}|\, |\Nmat + \Vmat^\top \Lmat\, \Vmat|$. Equating the two expressions yields Eq.~\eqref{eq:mdl}. The identity $|\Imat + \mathbf{A}\mathbf{B}| = |\Imat + \mathbf{B}\mathbf{A}|$ for any rectangular $\mathbf{A}, \mathbf{B}$ of compatible shape, used implicitly above, is a standard consequence of the same block-matrix construction \citep{GolubVanLoan2013}.

Applying Eq.~\eqref{eq:mdl} to the \cf factorisation $\Cmat = \Nmat + \Vmat^\top \Lmat\, \Vmat$ (Eq.~\ref{eq:C_factored}) gives
\begin{equation}
    |\Cmat| = |\Nmat|\, \left| \Imat_{\nmodes} + \Lmat\, \Mmat \right|,
    \qquad \Mmat \equiv \Vmat\, \Nmat^{-1}\, \Vmat^\top.
\end{equation}
The bracket can be rewritten using $\Lmat\, \Kmat = \Imat_{\nmodes} + \Lmat\, \Mmat$ with $\Kmat = \Lmat^{-1} + \Mmat$ (Eq.~\ref{eq:K_kernel}), so that $|\Imat_{\nmodes} + \Lmat\, \Mmat| = |\Lmat|\, |\Kmat|$. Substituting and taking logarithms recovers Eq.~\eqref{eq:logdet}:
\begin{equation}
    \ln|\Cmat| = \ln|\Nmat| + \ln|\Lmat| + \ln|\Kmat|.
\end{equation}
Every term on the right-hand side is a determinant of a matrix whose dimension does not exceed $\nmodes$. The pixel-space determinant $|\Cmat|$ is never formed.

\subsection{Numerical evaluation in \cf}
\label{app:logdet_numerical}

\paragraph{$\ln|\Nmat|$.} The noise covariance $\Nmat$ is independent of the cosmological parameters, so $\ln|\Nmat|$ is computed once at setup from a Cholesky factorisation $\Nmat = \mathbf{L}\,\mathbf{L}^\top$ as $\ln|\Nmat| = 2 \sum_i \ln \mathbf{L}_{ii}$, and the result is cached for the lifetime of the analysis.

\paragraph{$\ln|\Lmat|$.} The harmonic signal covariance $\Lmat$ is block-diagonal in $(\ell, m)$. The structure of each block depends on the field configuration:

\begin{itemize}
    \item \emph{Single spin-0 field.} $\Lmat_{\ell m} = C_\ell W_\ell$ is a scalar, and $\ln|\Lmat| = \sum_\ell (2\ell+1) \ln(C_\ell W_\ell)$ where $W_\ell$ is the combined beam and pixel-window factor (Sect.~\ref{sec:V_operator}).
    \item \emph{Single spin-2 field.} $\Lmat_{\ell m}$ is the $2 \times 2$ block of Eq.~\eqref{eq:Lambda_spin2}, mixing $EE$, $BB$, and $EB$. Its determinant is computed in closed form per $(\ell, m)$ and summed.
    \item \emph{Multi-field (e.g.\ $T,E,B$ joint).} $\Lmat_{\ell m}$ is the $3\times 3$ block of Eq.~\eqref{eq:Lambda_TEB}; we evaluate $\ln|\Lmat_{\ell m}|$ via a Cholesky factorisation per $(\ell, m)$. The per-block cost is negligible compared to the assembly of $\Kmat$.
\end{itemize}
Because $\Lmat$ enters Eq.~\eqref{eq:logdet} with a positive sign, no sign-tracking ambiguity arises here; the lemma is stated in terms of $\Lmat$ itself rather than $\Lmat^{-1}$.

\paragraph{$\ln|\Kmat|$.} The kernel $\Kmat$ is a dense $\nmodes \times \nmodes$ matrix that depends on $\Cl$ through $\Lmat$. We evaluate $\ln|\Kmat|$ from the same Cholesky factorisation of $\Kmat$ that is already required for the inversion in Eq.~\eqref{eq:smw}, at no additional asymptotic cost. When $\Kmat$ is reduced in dimension by the $\ell$-switching scheme of Sect.~\ref{sec:ell_switching}, the multipoles outside the inference window (i.e.\ $\ell \in [\lminsig, \lmin) \cup (\lmax, \lmaxsig]$) migrate into an effective noise term $\Neff$ whose determinant is absorbed into $\ln|\Nmat|$, leaving Eq.~\eqref{eq:logdet} formally unchanged with $\Nmat \to \Neff$ and $\Lmat,\Kmat$ restricted to $\ell \in [\lmin, \lmax]$.

\paragraph{Quadratic form.} The pixel-space quadratic form $\dvec^\top \Cmat^{-1} \dvec$ that appears alongside the log-determinant in the Gaussian likelihood is evaluated by direct application of Eq.~\eqref{eq:smw}:
\begin{equation}
    \dvec^\top \Cmat^{-1} \dvec
    \;=\; \dvec^\top \Nmat^{-1} \dvec
    \;-\; \left(\Vmat\,\Nmat^{-1}\,\dvec\right)^{\!\top}
          \Kmat^{-1}\,
          \left(\Vmat\,\Nmat^{-1}\,\dvec\right).
    \label{eq:quad_form_smw}
\end{equation}
The first term depends only on the data and the cached $\Nmat^{-1}$; the second involves a single back-substitution against the Cholesky factor of $\Kmat$, the same factorisation used for the log-determinant. Both terms scale as $\mathcal{O}(\nmodes^3)$ for the Cholesky factorisation and $\mathcal{O}(\nmodes^2)$ for the back-substitution; the pixel-space cost $\mathcal{O}(\npix^3)$ is avoided entirely.

\section{Algorithmic optimisations and implementation}
\label{app:performance}

The harmonic reformulation of Sect.~\ref{sec:harmonic} reduces the leading cost of QML estimation from $\mathcal{O}(\npix^3)$ to $\mathcal{O}(\nmodes^3)$, but several further optimisations are required for analyses to remain tractable at $\nside \gtrsim 32$. This appendix collects the algorithmic and implementation-level techniques used in \cf, with explicit complexity statements. None of the optimisations described here introduces an approximation: each is an algebraically exact restatement of an operation already defined in the main text.

\subsection{Sparse evaluation of Fisher traces}
\label{app:sparse_traces}

The dominant cost of the harmonic-space Fisher matrix is the evaluation of the $(\nspec\,\nell)^2$ traces in Eq.~\eqref{eq:fisher_harmonic}. A naive implementation that forms the dense matrix products $\Ctinv\Emat_\ell$ scales as $\mathcal{O}(\nell^2\,\nmodes^2)$. This dense evaluation is unnecessary because each per-multipole derivative $\Emat_\ell$ is extremely sparse: it has only $(2\ell+1)$ nonzero entries on the diagonal (one per real harmonic mode at multipole $\ell$). Cross-spectrum derivatives ($EB$, $TE$, $TB$) similarly carry $2(2\ell+1)$ nonzero off-diagonal entries connecting the corresponding mode blocks.

Storing each $\Emat_\ell$ in coordinate (COO) form as triplets $(r_\alpha, c_\alpha, v_\alpha)$, the trace in Eq.~\eqref{eq:fisher_harmonic} reduces to
\begin{equation}
    F_{\ell\ell'} = \frac{1}{2}\sum_{\alpha,\beta} v_\alpha\, v_\beta\, \Ctinv[c_\beta,\, r_\alpha]\, \Ctinv[c_\alpha,\, r_\beta],
    \label{eq:sparse_trace}
\end{equation}
where $\alpha$ runs over the nonzero entries of $\Emat_\ell$ and $\beta$ over those of $\Emat_{\ell'}$. The cost of this sub-block evaluation is $\mathcal{O}(N_\ell\, N_{\ell'})$, with $N_\ell$ the nonzero count of $\Emat_\ell$. Summed over the Fisher matrix, the total scaling becomes
\begin{equation}
    \mathcal{O}\!\left(\sum_{\ell,\ell'} (2\ell+1)(2\ell'+1)\right) = \mathcal{O}(\lmax^4),
    \label{eq:sparse_trace_scaling}
\end{equation}
to be compared with $\mathcal{O}(\lmax^6)$ in pixel space and with $\mathcal{O}(\nell^2\, \nmodes^2) \sim \mathcal{O}(\lmax^6)$ for the dense harmonic-space evaluation.\footnote{The $\nmodes^2$ factor in the dense evaluation absorbs the full $\sum_\ell (2\ell+1) = (\lmax+1)^2$ mode count, hence the same $\lmax^6$ scaling as the pixel-space algorithm despite the smaller prefactor.} A similar reduction applies to the noise bias of Eq.~\eqref{eq:noise_bias} and to the per-bin operators of Sect.~\ref{sec:binning}, where the binned derivative inherits the same union-of-supports structure.

\paragraph{Generality.} The optimisation is purely algebraic: it makes no assumption on the structure of $\Nmat$, $\Vmat$ or $\Lmat$, and applies identically to spin-0, spin-2 and multi-field analyses. For reference, the symbolic-trace technique used by {\tt ECLIPSE} \citep{BilbaoAhedo2021} achieves a comparable $\mathcal{O}(\lmax^4)$ scaling, but only under the assumption of diagonal pixel noise; Eq.~\eqref{eq:sparse_trace} retains the same algorithmic complexity for arbitrary noise covariances.

\paragraph{Empirical gain.} Measured on a single 48-core node, the trace stage of a $QU$ analysis at $\nside=16$ drops from $71\,\mathrm{s}$ to $0.3\,\mathrm{s}$ ($\sim$\,$240\times$); at $\nside=32$ the same stage drops from approximately $4400\,\mathrm{s}$ to $5\,\mathrm{s}$ ($\sim$\,$850\times$). After this optimisation the dominant runtime contributions become covariance I/O, signal-matrix construction and the SMW kernel inversion (see Sect.~\ref{app:performance_summary}).

\subsection{Field block-diagonal SMW kernel}
\label{app:field_blocks}

For multi-field analyses the SMW kernel $\Kmat$ acts on a vector of dimension $\nmodes^\mathrm{total} = \sum_g \nmodes^g$, summed over field groups (e.g., $T$ alone, $EB$ together). Whenever (i) no cross-spectrum is included between two field groups and (ii) the noise covariance has no off-diagonal block coupling them, both $\Lmat$ and $\Vmat\Nmat^{-1}\Vmat^\top$ become block-diagonal across those groups, and so does $\Kmat$. Inverting each block independently replaces an $\mathcal{O}((\nmodes^\mathrm{total})^3)$ factorisation with $\sum_g \mathcal{O}((\nmodes^g)^3)$.

For the common $TEB$ analysis with no $TE$ or $TB$ cross-spectra and uncorrelated $T$/$P$ noise, two blocks remain ($T$ and $EB$) of comparable size, giving roughly a $4\times$ speedup on $\Kmat$ inversion. \cf detects this structure automatically from the spectra list and the noise covariance topology --- no user flag is required --- and falls back to a single dense inversion when any cross-coupling is present, ensuring correctness in the general case.

\subsection{Stable evaluation of the SMW projected inverse}
\label{app:smw_stable}

The Sherman--Morrison--Woodbury identity used in Sect.~\ref{sec:smw} writes the projected inverse as a difference of two large matrices,
\begin{equation}
    \Vmat\,\Cmat^{-1}\,\Vmat^{\top} = \Mmat - \Mmat\,\Kmat^{-1}\,\Mmat,
    \qquad
    \Kmat = \Lmat^{-1} + \Mmat,
    \label{eq:smw_subtractive}
\end{equation}
with $\Mmat \equiv \Vmat\Nmat^{-1}\Vmat^{\top}$. In the cosmic-variance-limited regime, where the signal contribution to the covariance dominates the noise (e.g., temperature at low multipoles when the polarization sensitivity is sub-$\mu\mathrm{K}\cdot\mathrm{arcmin}$), one has $\Lmat \gg \Mmat^{-1}$, and the two matrices on the right-hand side of Eq.~\eqref{eq:smw_subtractive} become large and nearly equal: the difference is dominated by catastrophic cancellation in floating-point arithmetic and can produce spuriously negative entries on the diagonal of $\Vmat\Cmat^{-1}\Vmat^{\top}$. When that happens the binned Fisher matrix $\Fbm$ acquires negative eigenvalues, breaking the positive-definiteness assumed by Cholesky inversion and contaminating any downstream covariance estimate.

\cf avoids the subtraction entirely. The algebraic identity
\begin{equation}
    \Mmat - \Mmat\,\Kmat^{-1}\,\Mmat \;=\; \Mmat\,\Kmat^{-1}\,\Lmat^{-1} \;=\; \Mmat\,(\Imat + \Lmat\Mmat)^{-1}
    \label{eq:smw_stable}
\end{equation}
follows from $\Kmat = \Lmat^{-1}(\Imat + \Lmat\Mmat)$, so $\Vmat\Cmat^{-1}\Vmat^{\top}$ can be evaluated as one solve of the linear system $(\Imat + \Lmat\Mmat)^{\top}\,\mathbf{X}^{\top} = \Mmat^{\top}$. The matrix $(\Imat + \Lmat\Mmat)$ has eigenvalues $1 + \mathrm{eig}(\Lmat^{1/2}\Mmat\Lmat^{1/2}) \ge 1$ and is therefore well-conditioned independently of the signal-to-noise ratio. Two further QML quantities follow the same identity:
\begin{align}
    \Vmat\,\Cmat^{-1}\,\mathbf{d} &= (\Imat + \Mmat\Lmat)^{-1}\,\Vmat\,\Nmat^{-1}\,\mathbf{d}, \label{eq:smw_stable_data} \\
    \Amat \;\equiv\; \Imat - \Mmat\,\Kmat^{-1} &= (\Imat + \Mmat\Lmat)^{-1}, \label{eq:smw_stable_A}
\end{align}
where $\mathbf{d}$ is a data vector and $\Amat$ enters the noise-bias matrix $\mathrm{Cov}(\mathbf{w}\,|\,\mathrm{noise}) = \Amat\,\Tmat\,\Amat^{\top}$ used in the QML pipeline. Equations~(\ref{eq:smw_stable})--(\ref{eq:smw_stable_A}) involve the same matrix $(\Imat + \Lmat\Mmat)$ (or its transpose), so a single LU factorisation is reused for the projected inverse, the data weighting and the noise-bias matrix. The wall-time cost is comparable to the legacy SMW form (one LU in place of one Cholesky plus dense matrix products); the gain is numerical accuracy in the high-SNR regime, where the legacy form silently corrupts the Fisher matrix.

\subsection{Implementation notes}
\label{app:implementation}

The pixel-space noise covariance is loaded via tiled {\tt numpy} fancy indexing (unmasked-pixel index split into $\sim\!64$\,MiB tiles), so that the active-mask submatrix is assembled in $\mathcal{O}(\lceil \npix/N_\mathrm{tile}\rceil^2)$ tile iterations without a full-size intermediate; signal-covariance assembly and spin-2 $\Vmat$ construction are {\tt Numba} {\tt @njit} kernels paired with direct LAPACK bindings, leaving no critical path in the Python interpreter. The sparse derivatives $\{\Emat_\ell\}$ built during the Fisher computation are cached and reused by \qube\ at the power-spectrum stage at no extra setup cost. On multi-rank nodes the largest precomputed objects ($\Nmat$, $\Smat$, the SMW factors) are placed in MPI shared-memory windows, so that per-node footprint is set by the object sizes rather than by the rank count. Numerically, the $(2\ell+1)/(4\pi)$ prefactor is absorbed into the spherical-harmonic basis functions (eliminating a class of post-hoc normalisation errors at $\ell = 2$), Wigner-$d$ matrix elements use the Jacobi-derived three-term recurrence (Appendix~\ref{app:spin_harmonics}), Fisher decorrelation regularises eigenvalues below $10^{-12}$ of the maximum, and log-determinants flow through the matrix-determinant lemma (Appendix~\ref{app:smw_logdet}) on LAPACK Cholesky factors so that no large determinant is ever computed directly.

\subsection{Summary of complexity reductions}
\label{app:performance_summary}

Table~\ref{tab:complexity} summarises the asymptotic costs of the dominant stages of a \cf QML analysis, with and without the optimisations of this appendix.

\begin{table}[ht]
\caption{Asymptotic complexity of the dominant QML stages. ``Pixel-direct'' refers to a direct evaluation of Eqs.~(\ref{eq:fisher})--(\ref{eq:q_ell}) in pixel space, with the per-bandpower precompute $\Ctinv\Emat_b$ shared across the trace pairs. ``Harmonic (dense)'' is the SMW reformulation of Sect.~\ref{sec:harmonic} with naive trace evaluation. ``Harmonic (\cf)'' adds the optimisations of this appendix. The dimensions are: $\npix$ unmasked pixels, $\nmodes$ harmonic modes ($\sim\lmax^2$), $\nell = \lmax-\lmin+1$ multipoles, $\nbins$ bandpowers ($\nbins \le \nell$, with $\nbins = \nell$ in the unbinned case), $\nspec$ spectra. Big-$\mathcal{O}$ scaling is implicit throughout.}
\label{tab:complexity}
\vskip -3mm
\nointerlineskip
\setbox\tablebox=\vbox{
\halign{
#\hfil\tabskip 0.6em&
\hfil#\hfil\tabskip 0.6em&
\hfil#\hfil\tabskip 0.6em&
\hfil#\hfil\tabskip 0pt\cr
\noalign{\doubleline}
\omit\hfil Stage\hfil& \shortstack{Pixel-\\direct}& \shortstack{Harmonic\\(naive)}& \shortstack{Harmonic\\(\cf)}\cr
\noalign{\vskip 3pt\hrule\vskip 3pt}
$\Cmat^{-1}$ assembly  & $\npix^3$ & $\nmodes^3$ & $\nmodes^3$\,$^{\dagger}$\cr
Fisher matrix\,$^{\ddagger}$ & ${\nspec\,\nbins\,\npix^3 \atop +\,\nspec^2\,\nbins^2\,\npix^2}$ & $\nbins^2\,\nmodes^2$ & $\lmax^4$\cr
Quadratic estimates    & $\nbins\,\npix^2$ & $\nbins\,\nmodes$ & $\nmodes$\cr
\noalign{\vskip 3pt\hrule\vskip 3pt}}}
\endPlancktable
\tablenote{\dagger} Split into block sums when the kernel is field block-diagonal (Sect.~\ref{app:field_blocks}).\par
\tablenote{\ddagger} Pixel-direct cost is dominated by the $\nbins$ Cholesky-multiplies $\Ctinv\Emat_b$ when $\npix \gg \nbins$; for $\nbins \to \nell$ (unbinned) the leading term scales as $\nell\,\npix^3 \sim \lmax^7$ at fixed sky fraction.\par
\end{table}

For a representative $TEB$ analysis at $\nside=32$, $\lmax=64$, $\fsky\approx 0.7$, the combined effect of the optimisations in this appendix is to bring a single Fisher$\,+\,$Spectra evaluation from a regime where it is dominated by Fisher trace evaluation ($\sim$\,hours per rank) to one where it is dominated by setup cost (covariance I/O, signal matrix and SMW kernel construction; minutes per rank).

\subsection{Memory profile derivation and calibration}
\label{app:perf_memory_appendix}

\subsubsection{Persistent-state model}
\label{sec:perf_memory_model}

The persistent state retained by each computation basis through the Fisher and \texttt{Spectra} stages admits a closed-form size estimate in terms of the basis dimensions; the dominant transients are short enough not to set the lifetime peak. We collect the leading terms here so that a user can predict the per-rank footprint of a planned analysis before submission. All terms are dense double-precision buffers and carry an implicit factor of $8$\,bytes.

\paragraph{Harmonic path.} The persistent state retained from \texttt{basis\_setup} onwards is
\begin{equation}
    M_\mathrm{harm}^\mathrm{pers} =
    \underbrace{\npix^2}_{\mathbf{L}}
    + \underbrace{\nmodes \cdot \npix}_{\mathbf{V}\Nmat^{-1}}
    + \underbrace{\nmodes^2}_{\mathbf{V}\Nmat^{-1}\mathbf{V}^\top}
    + \underbrace{\nmodes^2}_{\mathbf{T},\;\mathrm{switch\;only}},
    \label{eq:budget_harmonic_pers}
\end{equation}
where $\mathbf{L}$ is the in-place Cholesky factor of $\Nmat$ (the explicit $\Nmat^{-1}$ is never formed), $\mathbf{V}\Nmat^{-1}$ and $\mathbf{V}\Nmat^{-1}\mathbf{V}^\top$ are the precomputed projectors of Sect.~\ref{sec:fisher_harmonic}, and $\mathbf{T}$ is the noise-bias kernel that decouples from $\mathbf{V}\Nmat^{-1}\mathbf{V}^\top$ only when $\ell$-switching (Sect.~\ref{sec:ell_switching}) is active. The \texttt{basis\_setup} transient peak adds the pixel projector $\Vmat$ itself ($\nmodes \cdot \npix$) and, when switching is active, the fixed-multipole signal contribution $S_\mathrm{fixed}$ ($\npix^2$) together with one $\nmodes \cdot \npix$ intermediate.

\paragraph{Pixel-direct path.} Without the harmonic projector and SMW kernel, the persistent state is dominated by $\npix^2$ buffers:
\begin{equation}
    M_\mathrm{pix}^\mathrm{pers} =
    \underbrace{\npix^2}_{\Cmat}
    + \underbrace{\npix^2}_{\Nmat\;\mathrm{(F\text{-}order\;copy)}}
    + \underbrace{\npix^2}_{S_\mathrm{fixed}\;\mathrm{(switch\;only)}}.
    \label{eq:budget_pixel_pers}
\end{equation}
The middle term reflects an unavoidable Fortran-ordered copy at basis-construction time; the third term is allocated only when the basis-construction range $\lmaxsig$ exceeds the inference window $\lmax$ (the implicit-switch regime), and is held in the allocator pool through \texttt{basis\_setup} exit even after its consumer dereferences it. The Fisher transient peak adds $\Cmat^{-1}$ together with the per-bin products $\{\Cmat^{-1}\,\partial \Cmat/\partial \theta_i\}$ that drive the trace loop, scaling as $n_\mathrm{params}\,\npix^2$ with $n_\mathrm{params} = \nspec\,\nbins$. The Fisher constructor exposes \texttt{cache\_derivatives = False} to suppress this cache at the cost of recomputing the products inside \texttt{Spectra}, recovering a $\sim\!2\times$ saving on the transient term; the default caches them in line with the throughput-oriented baseline of Sect.~\ref{sec:perf_scaling}.

The \texttt{qube-memory-budget} command-line tool exposes Eqs.~(\ref{eq:budget_harmonic_pers})--(\ref{eq:budget_pixel_pers}) and the corresponding transient terms, returning a per-stage predicted peak that a user can consult before requesting a node allocation.

\subsubsection{Measured profile and calibration}
\label{sec:perf_memory_calib}

Table~\ref{tab:memory_pixel} reports the pixel-direct path at $\fsky = 0.1$. Memory is dominated by the Fisher stage; Spectra reuses Fisher's setup and adds at most a few megabytes for the ten-simulation batch. At $\nside = 64$, $QU$ the per-bin derivative cache (the leading transient identified above) carries $\sim\!13$\,GiB; the remainder of the $31.1$\,GB peak is the persistent $\npix^2$ buffers of Eq.~\eqref{eq:budget_pixel_pers}, the Python interpreter baseline ($\sim\!13$\,GiB at the imports used here), and BLAS scratch.

\begin{table}
\caption{Peak RSS for the pixel-direct path at $\fsky = 0.1$ (single
rank, full thread budget). The larger of the two pipeline-stage peaks
is the node-sizing number.}
\label{tab:memory_pixel}
\vskip -3mm
\nointerlineskip
\setbox\tablebox=\vbox{
\newdimen\digitwidth \setbox0=\hbox{\rm 0} \digitwidth=\wd0
\catcode`*=\active \def*{\kern\digitwidth}
\halign{
\hfil#\hfil\tabskip 1.5em&
\hfil#\tabskip 1.5em&
\hfil#\tabskip 1.5em&
\hfil#\tabskip 1.5em&
\hfil#\tabskip 0pt\cr
\noalign{\doubleline}
\omit\hfil Field\hfil& \omit\hfil $\nside$\hfil& \omit\hfil $\npix$\hfil& \omit\hfil Fisher peak (GB)\hfil& \omit\hfil Spectra peak (GB)\hfil\cr
\noalign{\vskip 3pt\hrule\vskip 3pt}
$T$  & 16 & *312 & *0.4 & *0.4\cr
$QU$ & 16 & *624 & *0.6 & *0.6\cr
$T$  & 32 & 1200 & *0.7 & *0.7\cr
$QU$ & 32 & 2400 & *2.4 & *1.7\cr
$T$  & 64 & 4900 & *3.8 & *3.2\cr
$QU$ & 64 & 9800 & 31.1 & 19.8\cr
\noalign{\vskip 3pt\hrule\vskip 3pt}}}
\endPlancktable
\end{table}

The harmonic path has a different working set --- dominated by the projector $\mathbf{V}\Nmat^{-1}$ and the projected inverse $\mathbf{V}\Nmat^{-1}\mathbf{V}^\top$ of Eq.~\eqref{eq:budget_harmonic_pers}, with sparse-COO derivatives contributing negligibly. Table~\ref{tab:memory_harmonic} reports the harmonic profile at $\fsky = 0.5$, where the harmonic basis is the cost-optimal choice; peak RSS at this sky fraction is dominated by the $\nmodes \cdot \npix$ and $\nmodes^2$ terms of Eq.~\eqref{eq:budget_harmonic_pers}.

\begin{table}
\caption{Peak RSS for the harmonic path at $\fsky = 0.5$ (single rank, full thread budget); complementary to Table~\ref{tab:memory_pixel}, each reported at the sky fraction where its basis is the \texttt{auto} selector's cost-optimal choice.}
\label{tab:memory_harmonic}
\vskip -3mm
\nointerlineskip
\setbox\tablebox=\vbox{
\newdimen\digitwidth \setbox0=\hbox{\rm 0} \digitwidth=\wd0
\catcode`*=\active \def*{\kern\digitwidth}
\halign{
\hfil#\hfil\tabskip 1.5em&
\hfil#\tabskip 1.5em&
\hfil#\tabskip 1.5em&
\hfil#\tabskip 1.5em&
\hfil#\tabskip 0pt\cr
\noalign{\doubleline}
\omit\hfil Field\hfil& \omit\hfil $\nside$\hfil& \omit\hfil $\npix$\hfil& \omit\hfil $n_\mathrm{modes}$\hfil& \omit\hfil Fisher peak (GB)\hfil\cr
\noalign{\vskip 3pt\hrule\vskip 3pt}
$T$  & 16 & *1568 & 1085 & *0.9\cr
$QU$ & 16 & *3136 & 2170 & *1.6\cr
$T$  & 32 & *6208 & 4221 & *4.5\cr
$QU$ & 32 & 12416 & 8442 & 15.7\cr
\noalign{\vskip 3pt\hrule\vskip 3pt}}}
\endPlancktable
\end{table}

The model of Eqs.~(\ref{eq:budget_harmonic_pers})--(\ref{eq:budget_pixel_pers}) has been calibrated against \texttt{psutil}-instrumented production runs: on a $QU$, $\nside = 64$, $\lmaxsig = 256$, $\lmax = 128$, $\fsky \approx 0.6$ harmonic stress test (and on the pixel-direct $\nside = 64$, $QU$, $\fsky = 0.1$ configuration of Table~\ref{tab:memory_pixel}), persistent state is reproduced at the $\sim\!1\%$ level on both paths; transient peaks at $\sim\!0.5\%$ on the harmonic side and within $\sim\!6\%$ on the pixel-direct \texttt{Fisher} peak once allocator and BLAS-scratch slack are included. The pixel-direct \texttt{Spectra} stage is modelled as a ceiling rather than a tight estimate, on the conservative side appropriate for cluster sizing.


\section{Comparison with existing codes}
\label{app:comparison}

The QML estimator has been implemented several times over the past two decades, each implementation tailored to the science case at hand and to the computing resources available at the time. The role of this section is to position \cf within that landscape --- not to claim superior performance against any specific competitor. The correctness of \cf is established independently in Sect.~\ref{sec:fortran_validation} through element-by-element validation against the Planck low-$\ell$ Fortran reference implementation. What follows is a feature-scope comparison in the spirit of \citet{Alonso2019}, who introduced \texttt{NaMaster} as a unified pseudo-$C_\ell$ tool succeeding a generation of single-purpose codes. The QML landscape today is at a similar stage: each public code addresses a useful subset of the problem, and the gap that \cf seeks to fill is not a missing feature but the unification of features across a single, modern, extensible Python framework.

\subsection{The QML code landscape}
\label{sec:landscape}

We focus on the QML implementations that are publicly available, documented, and used in the recent CMB literature. We deliberately exclude internal or unpublished codes, and we do not attempt a historical survey; \citet{BilbaoAhedo2021} and \citet{Vanneste2018} provide complementary discussions of the algorithmic lineage.

\paragraph{The Planck low-$\ell$ reference.} The Fortran code used as the reference in Sect.~\ref{sec:fortran_validation}, distributed at \url{https://baltig.infn.it/cosmology_ferrara/lowell-likelihood-analysis} and used in \citet{Pagano2020} and the \texttt{simall} low-$\ell$ likelihood of \citet{PlanckV2020}, comprises three modules: template fitting for polarisation, QML power spectrum estimation supporting both auto- and cross-spectra, and a pixel-based Gaussian likelihood evaluated on a parameter grid. It supports general (non-diagonal) noise covariances and the full $T,Q,U$ joint analysis. Its main limitation is the implementation language --- Fortran with custom build chains --- and the absence of native multipole binning; both were appropriate for the production context but represent a barrier to entry today.

\paragraph{ECLIPSE \citep{BilbaoAhedo2021}.} A Fortran implementation distributed at \url{https://github.com/CosmoTool/ECLIPSE}, designed around the observation that the QML algebra simplifies dramatically when the noise is spatially uncorrelated. ECLIPSE supports the full intensity-and-polarisation analysis, both pixel- and harmonic-space implementations of the algebra, and provides a binned version of the estimator that recovers a regular Fisher matrix when small sky fractions would otherwise render the standard estimator singular. The code is parallelised for high-performance computing centres (144 cores in the published benchmarks). Because ECLIPSE assumes diagonal noise --- an assumption that allows several traces and matrix-vector products to be replaced by sub-block operations --- it is exceptionally fast for experiments whose noise can be approximated as diagonal in pixel space, but does not natively support the correlated noise covariances that arise generically once residual systematics, component-separation outputs, or cross-frequency correlations are propagated through the analysis pipeline, in both space- and ground-based experiments.

\paragraph{xQML \citep{Vanneste2018}.} A Python code (with OpenMP-accelerated C routines) distributed at \url{https://gitlab.in2p3.fr/xQML/xQML}, designed specifically for cross-spectrum estimation between two independent maps as a means of removing noise bias without requiring an analytical model of the noise. xQML supports the full $T,Q,U$ analysis, accepts dense per-map noise covariances $N_A$ and $N_B$, and provides a \texttt{Bins} class for native multipole binning --- a feature which directly inspired the binning module of \cf (see~Sect.~\ref{sec:binning})\footnote{The \cc \texttt{Bins} class is adapted from xQML and we acknowledge this lineage explicitly.}. It does not include a pixel-space likelihood module and works entirely in the pixel basis: the dense pixel-space covariance $\Smat + \Nmat$ is inverted directly, with no harmonic-basis (SMW-style) path.

\paragraph{QML-FAST \citep{Kvasiuk2025}.} A Python code (NumPy + Numba) distributed at \url{https://github.com/ykvasiuk/qmlfast}, designed to scale QML estimation to large numbers of correlated scalar fields, motivated by galaxy-clustering applications. QML-FAST inverts the dense pixel-space covariance directly and then uses the Legendre addition theorem to project the Fisher traces through spherical harmonics, evaluating $\mathrm{Tr}[\Cmat^{-1} \Pmat_{\ell_1} \Cmat^{-1} \Pmat_{\ell_2}]$ via precomputed $\Ctinv \equiv \Vmat \Cmat^{-1} \Vmat^T$ contractions; this is harmonic-space algebra at the trace level, not a Sherman--Morrison--Woodbury reformulation of the covariance itself, so we mark the corresponding row in Table~\ref{tab:feature_comparison} as partial. It supports native band-power Fisher computation via eigendecomposition of binned basis elements, and arbitrary rank-one decompositions beyond the spherical-harmonic default. The public examples and the released API are restricted to scalar (spin-0) fields: the repository contains no spin-weighted spherical harmonic or Wigner-$d$ machinery, and polarisation is not documented as a supported configuration.

\paragraph{BolPol \citep{Gruppuso2009}.} A Fortran QML code developed for early Planck and WMAP polarisation analyses. It treats the full $T,Q,U$ covariance as a single dense matrix, and is paired with a companion pixel-space likelihood. The code itself, to our knowledge, has never been publicly released, which limits its accessibility outside the original collaboration; we include it in the comparison because of its historical role in the QML literature, but the entries in Table~\ref{tab:feature_comparison} that we cannot verify from a public source are flagged with ``?''.

Each of these codes is well-suited to the analysis it was developed for, and the choice of which code to deploy in a given study is a function of that study's specific requirements.

\subsection{Feature comparison}
\label{sec:feature_comparison}

Table~\ref{tab:feature_comparison} compiles the capabilities of \cf alongside the codes discussed above. Entries marked ``y'' denote a documented, publicly available feature; ``--'' denotes an absent or undocumented feature; ``p'' denotes partial support (the feature is present but covers a subset of \cf's implementation, e.g.\ QML-FAST evaluates Fisher traces through spherical-harmonic projections of a pixel-space inverse covariance rather than representing and inverting the covariance in a harmonic basis via SMW); ``?'' marks claims that the published material does not unambiguously establish. The standard, deconvolved $\hat C_\ell = \Fmat^{-1} q$ output is implicit in every row of the table since every QML estimator produces it; the rows ``decorrelated bandpowers'' and ``window-convolved estimates'' track only the two \emph{additional} normalisation modes that \cf exposes (Sect.~\ref{sec:normalisation}). We have verified each entry against the most recent publicly available paper or repository at the time of writing; we caution the reader that the public versions of these codes evolve, and the table reflects a snapshot.

\begin{table*}
\caption{Feature matrix of the QML implementations discussed in Sect.~\ref{sec:landscape}. ``y'' available; ``--'' absent or undocumented; ``p'' partial support (a subset of the modes available in \cf); ``?'' unverified against the published source. The ``Planck low-$\ell$'' column refers to the \texttt{pse\_qml} + \texttt{parameter\_estimation} Fortran codes used in \citet{Pagano2020} and the \texttt{simall} low-$\ell$ likelihood of \citet{PlanckV2020}. ECLIPSE: \citet{BilbaoAhedo2021}. xQML: \citet{Vanneste2018}. QML-FAST: \citet{Kvasiuk2025}. BolPol: \citet{Gruppuso2009}. ``Cross (multi-spec, one map)'' denotes the joint estimation of all auto- and cross-spectra obtainable by combining the different fields (possibly with different spins) carried by a single observation --- e.g.\ the six $TT,EE,BB,TE,TB,EB$ spectra of a CMB $T,Q,U$ map, or the full set of auto- and cross-spectra among multiple correlated scalar fields treated as a single dataset; ``Cross (multi-map)'' denotes the cross-correlation of two independent maps for noise-bias removal. The two ``?'' cells in the BolPol column reflect the absence of a public release for that code: the corresponding capabilities cannot be verified independently from the published paper, as discussed in Sect.~\ref{sec:landscape}.}
\label{tab:feature_comparison}
\vskip -3mm
\nointerlineskip
\setbox\tablebox=\vbox{
\halign{
#\hfil\tabskip 1.5em&
\hfil#\hfil\tabskip 1em&
\hfil#\hfil\tabskip 1em&
\hfil#\hfil\tabskip 1em&
\hfil#\hfil\tabskip 1em&
\hfil#\hfil\tabskip 1em&
\hfil#\hfil\tabskip 0pt\cr
\noalign{\doubleline}
\omit\hfil Feature\hfil& \cf& Planck low-$\ell$& ECLIPSE& xQML& QML-FAST& BolPol\cr
\noalign{\vskip 3pt\hrule\vskip 3pt}
Spin-0 ($T$)                  & y & y & y & y & y & y\cr
Spin-2 ($Q,U$)                & y & y & y & y & -- & y\cr
Cross (multi-spec, one map)   & y & y & y & y & y & y\cr
Cross (multi-map)             & y & y & -- & y & y & ?\cr
General (non-diagonal) noise  & y & y & -- & y & y & y\cr
Harmonic basis (SMW)          & y & -- & y & -- & p & --\cr
Pixel basis                   & y & y & y & y & y & y\cr
Native multipole binning      & y & -- & y & y & y & --\cr
Per-component multipole floor & y & -- & -- & -- & -- & --\cr
Decorrelated bandpowers       & y & -- & -- & -- & -- & --\cr
Window-convolved estimates    & y & -- & -- & -- & -- & --\cr
Pixel-space likelihood        & y & y & -- & -- & -- & y\cr
MPI parallelisation           & y & y & y & -- & -- & ?\cr
Public release                & y & y & y & y & y & --\cr
Language                      & Python & Fortran & Fortran & Python+C & Python & Fortran\cr
\noalign{\vskip 3pt\hrule\vskip 3pt}}}
\endPlancktablewide
\end{table*}

Four rows in the table are populated only in the \cf column: the decorrelated bandpower output, the window-convolved estimates, the simultaneous availability of both a harmonic-basis SMW path and a direct pixel-space path behind a single abstract interface, and a per-component multipole floor that lets the spin-0 estimation range start at $\ell = 1$ (or, in principle, at $\ell = 0$) while spin-2 components stay at the representation-theoretic minimum $\ell = 2$ in the same joint analysis. ECLIPSE implements both pixel- and harmonic-space algorithms internally but exposes them as alternative computational strategies rather than user-selectable bases; xQML, QML-FAST, the Planck low-$\ell$ reference, and BolPol each commit to one basis only.

\subsection{Positioning}
\label{sec:positioning}

A direct head-to-head runtime comparison with ECLIPSE in its diagonal-noise regime --- the only configuration in which a like-for-like measurement is possible without porting one of the two codes outside its design assumptions --- requires running both codes at matching $\nside$, $\lmax$ and mask, on the same hardware, and is deferred to a dedicated companion note where the configuration can be controlled in full. The purpose of the present section is feature scope rather than runtime benchmarking.

\cf fills, for QML analyses, the same gap that \texttt{NaMaster} \citep{Alonso2019} filled for pseudo-$C_\ell$ analyses: a single public, validated, modular Python framework that does the standard analysis correctly out of the box, can be extended to non-standard configurations without forking the codebase, and complements existing single-purpose implementations within one community-maintained codebase. Distribution channels (PyPI package, GitHub repository, online documentation) are listed in the conclusions (Sect.~\ref{sec:conclusions}).


\section{Configuration and YAML reference}
\label{app:yaml}

This appendix documents how a \cf analysis is configured and run, and records the YAML schema consumed by every analysis through the \texttt{InputParams} class. It is intended as a quick-reference for readers who want to reproduce the validation runs of Sect.~\ref{sec:fortran_validation} without consulting the online documentation. Default templates for the four canonical field configurations ($T$-only, $QU$-only, $TEB$, $TQU$) are bundled with the \qube distribution as \texttt{<config>\_defaults.yaml} files and serve as starting points on top of which user keys are layered.

\subsection{Configuration and usage}
\label{sec:configuration}

All analysis parameters are specified through YAML configuration files, including the HEALPix resolution ($\nside$), the multipole-range keys ($\lminsig$, $\lmaxsig$, $\lmin$, $\lmax$; see below), field types and labels, file paths for masks, noise covariance matrices, input power spectra, and beam specifications. Default templates are bundled for the canonical $T$-only, $QU$-only, $TEB$, and $TQU$ configurations. The public driver classes \texttt{Fisher}, \texttt{Spectra} (\qube) and \texttt{PICSLike} (\pics) all take a single YAML path on construction and a \texttt{run()} call to execute; \texttt{Spectra} accepts a precomputed \texttt{Fisher} instance to reuse cached derivatives and basis state. MPI parallelisation is transparent: the same script runs on a single core or across multiple processes via \texttt{mpirun}, with the workload distribution handled internally. The full API is documented online (see Sect.~\ref{sec:conclusions}).

\subsection{Top-level keys}
\label{app:yaml_top}

Table~\ref{tab:yaml_keys} lists the YAML keys recognised by \texttt{InputParams}, grouped by purpose. Required keys for a non-trivial analysis are flagged $\dagger$; the remaining keys default to a sensible value or are required only in specific modes (e.g., \texttt{covmatfile2} is required only when \texttt{do\_cross} is \texttt{true}). Computation-basis selection (harmonic vs.\ pixel) is performed at the API level rather than through YAML --- see Sect.~\ref{sec:harmonic_basis_code} --- and is therefore not listed here.

The field-label inputs accept a compact shorthand that is automatically expanded: a label string \texttt{"QU"} is interpreted as \texttt{["Q","U"]}, and underscore-separated combinations such as \texttt{"T1\_T2"} expand to \texttt{["T1","T2"]}. The \texttt{physical\_labels} key, when present, decouples the labels used for input-map FITS files from the labels used for spectra; this is needed when the maps are stored as $T,Q,U$ but spectra are estimated in $T,E,B$.

The multipole layout is controlled by four keys with an explicit constraint chain $\max(\texttt{lmin\_signal}) \le \texttt{lmin} \le \texttt{lmax} \le \texttt{lmax\_signal}$, and applies uniformly to both the QML estimator (\qube) and the pixel-space likelihood (\pics) since they share the same \texttt{Core} setup. The pair $(\texttt{lmin\_signal}, \texttt{lmax\_signal})$ delimits the \emph{signal-cov band} --- the range of multipoles represented in the basis ($\Vmat$, $\Lmat$, $\Smat$). The pair $(\texttt{lmin}, \texttt{lmax})$ selects the \emph{estimation/inference window} within that band: in \qube it is the range over which the QML quadratic estimator solves for $\hat C_\ell$, and in \pics it is the range over which the model $C_\ell$ are varied across the parameter grid. Multipoles inside the signal-cov band but outside this window are held fixed at the fiducial spectrum (read from \texttt{fiducialfile}) and contribute through $\Smat_\mathrm{fixed}$ on both the low-$\ell$ and high-$\ell$ side. The window $[\texttt{lmin}, \texttt{lmax}]$ corresponds to the symbols $[\lmin, \lmax]$ used in the formalism (Sect.~\ref{sec:ell_switching}); the basis range $[\texttt{lmin\_signal}, \texttt{lmax\_signal}]$ corresponds to $[\lminsig, \lmaxsig]$; the high-$\ell$ side recovers the original $\ell$-switching scheme as a special case. Defaults are \texttt{lmin\_signal=2}, \texttt{lmax\_signal=4*nside}, \texttt{lmin=2}, \texttt{lmax=lmax\_signal}, reproducing the behaviour $\ell \in [2, 4\nside]$ of analyses that pre-date this API.

The key \texttt{lmin\_signal} accepts either a scalar (broadcast to every component) or a per-component list, with the per-component constraint $\texttt{lmin\_signal}[i] \ge |\texttt{spins}[i]|$ enforcing the representation-theoretic floor. Setting \texttt{lmin\_signal=[1, 2]} on a \texttt{spins=[0, 2]} configuration enables direct estimation of the temperature dipole jointly with the polarisation spectra; setting \texttt{lmin\_signal=0} on a designated spin-0 component is the natural configuration for foreground-template marginalisation.

The conventions \texttt{input\_convention} and \texttt{output\_convention} control whether the input $C_\ell$ table and the output spectra are interpreted as $C_\ell$ or $D_\ell \equiv \ell(\ell+1) C_\ell / 2\pi$. Both default to \texttt{"Cl"}; users supplying CAMB or CLASS $D_\ell$ tables should override \texttt{input\_convention} to \texttt{"Dl"}.

\begin{table}
\caption{YAML keys recognised by \texttt{InputParams}, grouped by purpose.
Required keys are marked $\dagger$. ``Path'' refers to a file path resolved
relative to the YAML file's directory.}
\label{tab:yaml_keys}
\vskip -3mm
\nointerlineskip
\footnotesize
\setbox\tablebox=\vbox{
\halign{
#\hfil\tabskip 0.6em&
#\hfil\tabskip 0.6em&
#\hfil\tabskip 0pt\cr
\noalign{\doubleline}
\omit\hfil Key\hfil& \omit\hfil Type\hfil& \omit\hfil Role\hfil\cr
\noalign{\vskip 3pt\hrule\vskip 3pt}
\noalign{\vskip 2pt}\multispan3 \textit{Field configuration}\hfil\cr\noalign{\vskip 2pt}
\texttt{nside}$^\dagger$        & int       & HEALPix resolution\cr
\texttt{spins}$^\dagger$        & list[int] & per-field spin (0 or 2)\cr
\texttt{labels}$^\dagger$       & list[str] & spectrum labels\cr
\texttt{physical\_labels}       & list[str] & input-map labels\cr
\texttt{ordering}               & str       & {\tt RING}/{\tt NESTED}\cr
\noalign{\vskip 3pt\hrule\vskip 3pt}
\noalign{\vskip 2pt}\multispan3 \textit{Inputs}\hfil\cr\noalign{\vskip 2pt}
\texttt{inputclfile}            & path      & input $C_\ell$/$D_\ell$ table\cr
\texttt{maskfile}               & path      & analysis mask FITS\cr
\texttt{covmatfile1}            & path      & noise covariance, map 1\cr
\texttt{covmatfile2}            & path      & noise covariance, map 2\cr
\texttt{inputmapfile1}          & path      & input map FITS, map 1\cr
\texttt{inputmapfile2}          & path      & input map FITS, map 2\cr
\texttt{fiducialfile}           & path      & fiducial spectra (switching)\cr
\texttt{input\_convention}      & str       & {\tt Cl}/{\tt Dl}\cr
\noalign{\vskip 3pt\hrule\vskip 3pt}
\noalign{\vskip 2pt}\multispan3 \textit{Beam and pixel window}\hfil\cr\noalign{\vskip 2pt}
\texttt{smoothing\_type}        & str       & $\vcenter{\hbox{{\tt cosine\_legacy}/{\tt cosine\_npipe}/}\hbox{{\tt gaussian}/{\tt file}}}$\cr
\texttt{beam\_file}             & path      & $B_\ell$ FITS\cr
\texttt{fwhmarcmin}             & float     & Gaussian beam FWHM\cr
\texttt{apply\_pixwin}          & bool      & include $p_\ell$\cr
\texttt{smooth\_pol}            & bool      & smooth polarisation\cr
\noalign{\vskip 3pt\hrule\vskip 3pt}
\noalign{\vskip 2pt}\multispan3 \textit{Estimator and noise}\hfil\cr\noalign{\vskip 2pt}
\texttt{do\_cross}              & bool      & cross-spectrum mode\cr
\texttt{remove\_nb}             & bool      & subtract noise bias\cr
\texttt{calibration}            & float     & amplitude rescaling\cr
\texttt{load\_inverted}         & bool      & reuse cached $\Cmat^{-1}$\cr
\texttt{load\_reduced}          & bool      & noise cov.\ pre-reduced\cr
\noalign{\vskip 3pt\hrule\vskip 3pt}
\noalign{\vskip 2pt}\multispan3 \textit{Binning}\hfil\cr\noalign{\vskip 2pt}
\texttt{delta\_ell}             & int       & uniform bin width (default 1)\cr
\texttt{bin\_lmins}             & list[int] & explicit bin lower edges\cr
\texttt{bin\_lmaxs}             & list[int] & explicit bin upper edges\cr
\noalign{\vskip 3pt\hrule\vskip 3pt}
\noalign{\vskip 2pt}\multispan3 \textit{Multipole-range API}\hfil\cr\noalign{\vskip 2pt}
\texttt{lmin\_signal}           & int/list  & per-component floor ($\lminsig$)\cr
\texttt{lmax\_signal}           & int       & signal-cov ceiling ($\lmaxsig$)\cr
\texttt{lmin}                   & int       & estimation lower ($\lmin$)\cr
\texttt{lmax}$^\dagger$         & int       & estimation upper ($\lmax$)\cr
\noalign{\vskip 3pt\hrule\vskip 3pt}
\noalign{\vskip 2pt}\multispan3 \textit{Simulations and parameter grids}\hfil\cr\noalign{\vskip 2pt}
\texttt{nsims}                  & int       & number of simulations\cr
\texttt{root\_dir}              & path      & theory-spectra root dir\cr
\texttt{root\_filename}         & str       & theory-spectra template\cr
\texttt{parameters}             & dict      & cosmological grid\cr
\noalign{\vskip 3pt\hrule\vskip 3pt}
\noalign{\vskip 2pt}\multispan3 \textit{Outputs}\hfil\cr\noalign{\vskip 2pt}
\texttt{outinvcovmatfile1,2}    & path      & output $\Cmat^{-1}$\cr
\texttt{outnoisecovmat1,2}      & path      & output reduced $\Nmat$\cr
\texttt{outfilefisher}          & path      & output Fisher matrix\cr
\texttt{outcovmatfile}          & path      & output $C_\ell$ covariance\cr
\texttt{outerrfile}             & path      & output spectrum errors\cr
\texttt{output\_geometry\_file} & path      & output pixel geometry\cr
\texttt{output\_convention}     & str       & {\tt Cl}/{\tt Dl}\cr
\noalign{\vskip 3pt\hrule\vskip 3pt}
\noalign{\vskip 2pt}\multispan3 \textit{Logging}\hfil\cr\noalign{\vskip 2pt}
\texttt{feedback}               & int       & verbosity level\cr
\texttt{log\_file}              & path      & log destination\cr
\texttt{log\_format}            & str       & {\tt scientific}/{\tt plain}\cr
\texttt{log\_timing}            & bool      & include timing diagnostics\cr
\noalign{\vskip 3pt\hrule\vskip 3pt}}}
\endPlancktable
\end{table}

The two cosine \texttt{smoothing\_type} options expose the same raised-cosine apodising kernel $b_\ell = \tfrac{1}{2}\!\left[1 + \cos\!\left(\pi\,(\ell-\ell_1)/(\ell_2-\ell_1)\right)\right]$ for $\ell_1 < \ell \le \ell_2$ at two published parameter choices: \texttt{cosine\_legacy} uses $\ell_1 = \nside$ \citep{Benabed2009,Aghanim2020likelihood} while \texttt{cosine\_npipe} uses $\ell_1 = 1$ \citep{Akrami2020npipe}; both adopt $\ell_2 = 3\nside$.

\subsection{Worked example}
\label{app:yaml_example}

The $B$-mode validation of Sect.~\ref{sec:fortran_validation} runs from the following configuration (paths are relative to the YAML file's directory). Setting \texttt{lmax\_signal} equal to \texttt{lmax} pins the basis to the Fortran reference, which has no signal beyond its own \texttt{lmax}; the default \texttt{lmax\_signal}\,$=\,4\nside$ otherwise routes the fiducial high-$\ell$ signal through $\Smat_\mathrm{fixed}$.

\begin{verbatim}
nside: 8
lmin_signal: 2
lmax_signal: 24
lmin: 2
lmax: 24
spins:  [0]
labels: ["B"]

maskfile:    inputs/mask.fits
inputclfile: inputs/dls.txt
input_convention: Dl
covmatfile1: inputs/ncov.bin
ordering:    RING

smoothing_type: file
beam_file:    inputs/beam.fits
apply_pixwin: false

delta_ell: 1
remove_nb: true
feedback:  1
\end{verbatim}

This file reproduces the \cf side of the validation in Sect.~\ref{sec:fortran_validation}; the pixel-likelihood validation reuses it, with the parameter grid and simulation paths supplied through the \texttt{ParameterGrid} API. Direct temperature-dipole estimation is enabled by \texttt{spins=[0]}, \texttt{lmin\_signal: [1]}, \texttt{lmin: 1}; a joint $T,Q,U$ analysis with the dipole estimated alongside the polarisation spectra uses \texttt{spins=[0, 2]}, \texttt{lmin\_signal: [1, 2]}.

\end{appendix}
\end{document}